# Coding for Parallel Channels: Gallager Bounds for Binary Linear Codes with Applications to Repeat-Accumulate Codes and Variations

Igal Sason      Idan Goldenberg

Technion – Israel Institute of Technology
Department of Electrical Engineering
Haifa 32000, Israel
{sason@ee, idang@tx}.technion.ac.il

January 16, 2018

## Abstract

This paper is focused on the performance analysis of binary linear block codes (or ensembles) whose transmission takes place over independent and memoryless parallel channels. New upper bounds on the maximum-likelihood (ML) decoding error probability are derived. These bounds are applied to various ensembles of turbo-like codes, focusing especially on repeat-accumulate codes and their recent variations which possess low encoding and decoding complexity and exhibit remarkable performance under iterative decoding. The framework of the second version of the Duman and Salehi (DS2) bounds is generalized to the case of parallel channels, along with the derivation of their optimized tilting measures. The connection between the generalized DS2 and the 1961 Gallager bounds, addressed by Divsalar and by Sason and Shamai for a single channel, is explored in the case of an arbitrary number of independent parallel channels. The generalization of the DS2 bound for parallel channels enables to re-derive specific bounds which were originally derived by Liu et al. as special cases of the Gallager bound. In the asymptotic case where we let the block length tend to infinity, the new bounds are used to obtain improved inner bounds on the attainable channel regions under ML decoding. The tightness of the new bounds for independent parallel channels is exemplified for structured ensembles of turbo-like codes. The improved bounds with their optimized tilting measures show, irrespectively of the block length of the codes, an improvement over the union bound and other previously reported bounds for independent parallel channels; this improvement is especially pronounced for moderate to large block lengths. However, in some cases, the new bounds under ML decoding happen to be a bit pessimistic as compared to computer simulations of sub-optimal iterative decoding, thus indicating that there is room for further improvement.

*Index Terms:* Block codes, distance spectrum, input-output weight enumerator (IOWE), linear codes, maximum-likelihood (ML) decoding, memoryless binary-input output-symmetric (MBIOS) channels, parallel channels, repeat-accumulate (RA) codes.

# 1 Introduction

We analyze the error performance of linear codes where the codewords are partitioned into several disjoint subsets, and the bits in each subset are transmitted over a certain communication channel. This scenario can be viewed as the transmission of information over a set of parallel channels, where every bit is assigned to one of these channels. Code partitioning is employed in transmission over block-fading channels (for performance bounds of coded communication systems over block-fading channels, see, e.g., [14, 35]), rate-compatible puncturing of turbo-like codes (see, e.g., [15, 31]), incremental redundancy retransmission schemes, cooperative coding, multi-carrier signaling (for performance bounds of coded orthogonal-frequency division multiplexing (OFDM) systems, see e.g., [34]), and other applications.

In his thesis [11], Ebert considered the problem of communicating over parallel discrete-time channels, disturbed by arbitrary and independent additive Gaussian noises, where a total power constraint is imposed upon the channel inputs. He found explicit upper and lower bounds on the ML decoding error probability, which decrease exponentially with block length. The exponents of the upper and lower bounds coincide for zero rate and for rates between the critical rate ($R_{\text{crit}}$) and capacity. The results were also shown to be applicable to colored Gaussian noise channels with an average power constraint on the channel.

Tight analytical bounds serve as a potent tool for assessing the performance of modern error-correction schemes, both for the case of finite block length and in the asymptotic case where the block length tends to infinity. In the setting of a single communication channel and by letting the block length tend to infinity, these bounds are applied in order to obtain a noise threshold which indicates the minimum channel conditions necessary for reliable communication. When generalizing the bounds to the scenario of independent parallel channels, this threshold is transformed into a multi-dimensional barrier within the space of the joint parallel-channel transition probabilities, dividing the space into attainable and non-attainable channel regions.

In [21], Liu et al. derive upper bounds on the ML decoding error probability of structured ensembles of codes whose transmission takes place over (independent) parallel channels. When generalizing an upper bound to the case of independent parallel channels, it is desirable to have the resulting bound expressed in terms of basic features of the code (or ensemble of codes), such as the distance spectrum. The inherent asymmetry of the parallel-channel setting poses a difficulty for the analysis, as different symbols of the codeword suffer varying degrees of degradation through the different parallel channels. This difficulty was circumvented in [21] by introducing a random mapper, i.e., a device which randomly and independently assigns symbols to the different channels according to a certain a-priori probability distribution. As a result of this randomization, Liu et al. derived upper bounds on the ML decoding error probability which solely depend on the weight enumerator of the overall code, instead of a specific split weight enumerator which follows from the partitioning of a codeword into several subsets of bits and the individual transmission of these subsets over different channels. The analysis in [21] modifies the 1961 Gallager-Fano bound from [12, Chapter 3] and adapts this bounding technique for communication over parallel channels. As special cases of this modified bound, a generalization of the union-Bhattacharyya bound, the Shulman-Feder bound [33], simplified sphere bound [8], and a combination of the two former bounds are derived for parallel channels. The upper bounds on the ML decoding error probability are applied to ensembles of codes defined on graphs (e.g., uniformly interleaved repeat-accumulate codes and turbo codes). The comparison in [21] between upper bounds under ML decoding and computer simulations of the performance of such ensembles under iterative decoding shows a good match in several cases. For a given ensemble of codes and a given codeword-symbol to channel assignment rule, a reliable channel region is defined as the closure of the set of parallel-channel transition probabilities for which the decoding error probability vanishes as the codeword length goes to infinity. The upper bounds on the block error probability derived in [21] enable to derive achievable regions for ensuring reliable communications under ML decoding.

Using the approach of the random mapper by Liu et al. [21], we derive a parallel-channel generalization of the DS2 bound [9, 29, 32] and re-examine, for the case of parallel channels, the well-known relations between this bound and the 1961 Gallager bound which exist for the single channel case [8, 32]. In this respect, it is shown in this paper that the generalization of the DS2 bound for independent parallel channels is not necessarily tighter than the corresponding generalization of the 1961 Gallager bound, as opposed to the case where the communication takes place over a single memoryless binary-input output-symmetric (MBIOS) channel.

The new bounds are used to obtain inner bounds on the boundary of the channel regions which are asymptotically (in the limit where we let the block length tend to infinity) attainable under ML decoding, and the results improve on those recently reported in [21]. The generalization of the DS2 bound for parallel channels enables to reproduce special instances which were originally derived by Liu et al. [21]. The tightness of these bounds for independent parallel channels is exemplified for structured ensembles of turbo-like codes, and the boundary of the improved attainable channel regions is compared with previously reported regions for Gaussian parallel channels, and shows significant improvement due the optimization of the tilting measures which are involved in the computation of the generalized DS2 and 1961 Gallager bounds for parallel channels.

The remainder of the paper is organized as follows. The system model is presented in Section 2, as well as preliminary material related to our discussion. In Section 3, we generalize the DS2 bound for independent parallel channels. Section 4 presents the 1961 Gallager bound from [21], and considers its connection to the generalized DS2 bound, along with the optimization of its tilting measures. Section 5 presents some special cases of these upper bounds which are obtained as particular cases of the generalized bounds in Sections 3 and 4. Attainable channel regions are derived in Section 6. Inner bounds on attainable channel regions for various ensembles of turbo-like codes and performance bounds for moderate block lengths are exemplified in Section 7. Finally, Section 8 concludes the paper. For a comprehensive tutorial paper on performance bounds of linear codes under ML decoding, the reader is referred to [29].

## 2 Preliminaries

In this section, we state the assumptions on which our analysis is based. We also introduce notation and preliminary material related to the performance analysis of binary linear codes whose transmission takes place over parallel channels.

### 2.1 System Model

We consider the case where the communication model consists of a parallel concatenation of $J$ statistically independent MBIOS channels, as shown in Fig. 1.

Using an error-correcting linear code $\mathcal{C}$ of size $M = 2^k$, the encoder selects a codeword $\underline{x}^m$ ($m = 0, 1, \ldots, M - 1$) to be transmitted, where all codewords are assumed to be selected with equal probability ($\frac{1}{M}$). Each codeword consists of $n$ symbols and the coding rate is defined as $R \triangleq \frac{\log_2 M}{n} = \frac{k}{n}$; this setting is referred to as using an $(n, k)$ code. The channel mapper selects for each coded symbol one of $J$ channels through which it is transmitted. The $j$-th channel component has a transition probability $p(y|x;j)$. The considered model assumes that the channel encoder performs its operation without prior knowledge of the specific mapping of the bits to the parallel channels. While in reality, the choice of the specific mapping is subject to the levels of importance of different coded bits, the considered model assumes for the sake of analysis that this mapping is random and independent of the coded bits. This assumption enables to average over all possible mappings, though suitable choices of mappings for the coded bits are expected to perform better than the average.

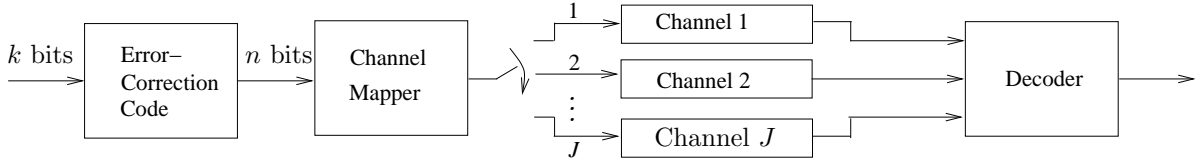

Figure 1: System model of parallel channels. A random mapper is assumed where every bit is assigned to one of the $J$ channels; a bit is assigned to the $j^{\text{th}}$ channel independently of the other bits and with probability $\alpha_j$ (where $\sum_{j=1}^{J} \alpha_j = 1$).

The received vector $\underline{y}$ is maximum-likelihood (ML) decoded at the receiver when the specific channel mapper is known at the receiver. While this broad setting gives rise to very general coding, mapping and decoding schemes, we will focus on the case where the input alphabet is binary, i.e., $x \in \{-1, 1\}$ (where zero and one are mapped to $+1$ and $-1$, respectively). The output alphabet is real, and may be either finite or continuous. By its definition, the mapping device divides the set of indices $\{1, \ldots, n\}$ into $J$ disjoint subsets $\mathcal{I}(j)$ for $j = 1, \ldots, J$, and transmits all the bits whose indices are included in the subset $\mathcal{I}(j)$ through the $j$-th channel. We will see in the next section that for a fixed channel mapping device (i.e., for given sets $\mathcal{I}(j)$), the problem of upper-bounding the ML decoding error probability is exceedingly difficult. In order to circumvent this difficulty, a probabilistic mapping device was introduced in [21] which uses a random assignment of the bits to the $J$ parallel channels; this random mapper takes a symbol and assigns it to channel $j$ with probability $\alpha_j$. This assignment is independent of that of other symbols, and by definition, the equality $\sum_{j=1}^{J} \alpha_j = 1$ follows. This approach enables in [21] the derivation of an upper bound for the parallel channels which is averaged over all possible channel assignments, and the bound can be calculated in terms of the distance spectrum of the code (or ensemble). Another benefit of the random mapping approach is that it naturally accommodates for practical settings where one is faced with parallel channels having different capacities.

### 2.2 Capacity Limit and Cutoff Rate of Parallel MBIOS Channels

We consider here the capacity and cutoff rate of independent parallel MBIOS channels. These information-theoretic quantities serve as a benchmark for assessing the gap under optimal ML decoding between the achievable channel regions of various ensembles of codes and the achievable channel region which corresponds to the Shannon capacity limit. It is also useful for providing a quantitative measure for the asymptotic performance of various ensembles as compared to the achievable channel region which corresponds to the cutoff rate of the considered parallel channels.

### 2.2.1 Cutoff Rate

The cutoff rate of an MBIOS channel is given by

$$R_0 = 1 - \log_2(1+\gamma) \tag{1}$$

where $\gamma$ is the Bhattacharyya constant, i.e.,

$$\gamma \triangleq \sum_y \sqrt{p(y|0)p(y|1)}. \tag{2}$$

Clearly, for continuous-output channels, the sum in the RHS of (2) is replaced by an integral.

For parallel MBIOS channels where every bit is assumed to be independently and randomly assigned to one of $J$ channels with a-priori probability $\alpha_j$ (where $\sum_{j=1}^{J} \alpha_j = 1$), the Bhattacharyya constant of the resulting channel is equal to the weighted sum of the Bhattacharyya constants of these individual channels, i.e.,

$$\gamma = \sum_{J=1}^{J} \alpha_j \sum_y \sqrt{p(y|0;j)p(y|1;j)}. \tag{3}$$

Consider a set of $J$ parallel Gaussian channels characterized by the transition probabilities

$$\begin{aligned} p(y|0;j) &= \frac{1}{\sqrt{2\pi}} e^{-\frac{(y+\nu_j)^2}{2}} \\ p(y|1;j) &= \frac{1}{\sqrt{2\pi}} e^{-\frac{(y-\nu_j)^2}{2}} \end{aligned} \tag{4}$$

$$-\infty < y < \infty, \; j = 1, \ldots, J$$

where

$$\nu_j \triangleq R \left(\frac{E_b}{N_0}\right)_j \tag{5}$$

and $\left(\frac{E_b}{N_0}\right)_j$ is the energy per information bit to the one-sided spectral noise density of the $j$-th channel. In this case, the Bhattacharyya constant is given by

$$\gamma = \sum_{j=1}^{J} \alpha_j e^{-\nu_j} \tag{6}$$

where $\nu_j$ is introduced in (5). From (1) and (6), the cutoff rate of $J$ parallel binary-input AWGN channels is given by

$$R_0 = 1 - \log_2 \left(1 + \sum_{j=1}^{J} \alpha_j e^{-R\left(\frac{E_b}{N_0}\right)_j}\right) \text{ bits per channel use.} \tag{7}$$

Consider the case of $J = 2$ parallel binary-input AWGN channels. Given the value of $\left(\frac{E_b}{N_0}\right)_1$, and the code rate $R$ (in bits per channel use), it is possible to calculate the value of $\left(\frac{E_b}{N_0}\right)_2$ of the second channel which corresponds to the cutoff rate. To this end, we set $R_0$ in the LHS of (7) to $R$. Solving this equation gives

$$\left(\frac{E_b}{N_0}\right)_2 = -\frac{1}{R} \ln \left(\frac{2^{1-R} - 1 - \alpha_1 e^{-R\left(\frac{E_b}{N_0}\right)_1}}{\alpha_2}\right). \tag{8}$$

### 2.2.2 Capacity Limit

Let $C_j$ designate the capacity (in bits per channel use) of the $j$-th MBIOS channel the set of $J$ parallel MBIOS channels. Clearly, by symmetry considerations, the capacity-achieving input distribution for all these channels is $\underline{q} = \left(\frac{1}{2}, \frac{1}{2}\right)$. The capacity of the $J$ parallel channels where each bit is randomly and independently assigned to the $j$-th channel with probability $\alpha_j$ is therefore given by

$$C = \sum_{j=1}^{J} \alpha_j C_j. \qquad (9)$$

For the case of $J$ parallel binary-input AWGN channels

$$C_j = 1 - \frac{1}{\sqrt{2\pi}\ln(2)} \int_{-\infty}^{\infty} e^{-\frac{(y-\beta_j)^2}{2}} \ln\left(1 + e^{-2\beta_j y}\right) dy \quad \text{bits per channel use} \qquad (10)$$

where $\beta_j \triangleq \sqrt{\frac{2E_s}{N_0}}$.

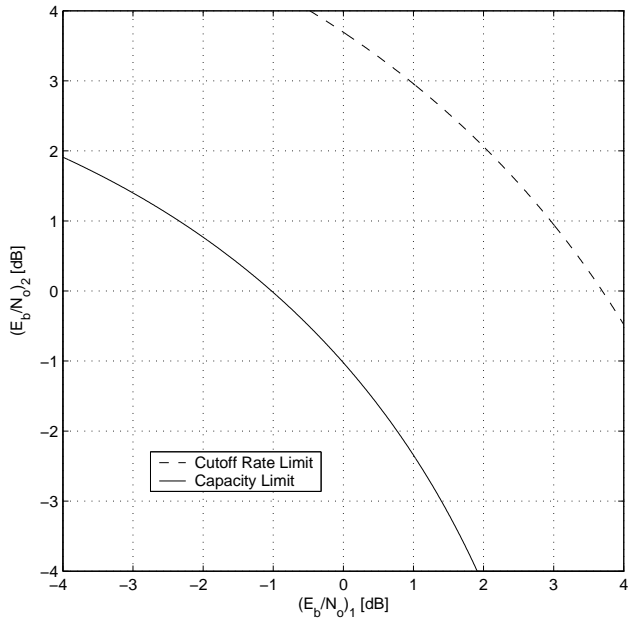

Figure 2: Attainable channel regions for two parallel binary-input AWGN channels, as determined by the cutoff rate and the capacity limit, referring to a code rate of one-third bits per channel use. It is assumed that each bit is randomly and independently assigned to one of these channels with equal probability (i.e., $\alpha_1 = \alpha_2 = \frac{1}{2}$).

In order to simplify the numerical computation of the capacity, one can express each integral in (10) as a sum of two integrals from 0 to $\infty$, and use the power series expansion of the logarithmic function; this gives an infinite power series with alternating signs. Using the Euler transform to expedite the convergence rate of these infinite sums, gives the following alternative expression:

$$C_j = 1 - \frac{1}{\ln(2)} \left[\frac{2\beta e^{-\frac{\beta_j^2}{2}}}{\sqrt{2\pi}} - (2\beta_j^2 - 1)Q(\beta_j) + \sum_{k=0}^{\infty} \frac{(-1)^k \cdot \Delta^k a_0(j)}{2^{k+1}}\right], \quad j = 1, \ldots, J \qquad (11)$$

where

$$\Delta^k a_0(j) \triangleq \frac{1}{2} e^{-\frac{\beta_j^2}{2}} \sum_{m=0}^{k} \left\{ \frac{(-1)^m}{(k-m+1)(k-m+2)} \binom{k}{m} \operatorname{erfcx}\left(\frac{(2k-2m+3)\beta_j}{\sqrt{2}}\right) \right\}$$

and

$$\operatorname{erfcx}(x) \triangleq 2e^{x^2} Q(\sqrt{2}x)$$

(note that $\operatorname{erfcx}(x) \approx \frac{1}{\sqrt{\pi}} \cdot \frac{1}{x}$ for large values of $x$). The infinite sum in (11) converges exponentially fast with $k$, and the summation of its first 30 terms happens to give very accurate result irrespective of the value of $\beta_j$.

Consider again the case of $J = 2$ parallel binary-input AWGN channels. Given the value of $\left(\frac{E_b}{N_0}\right)_1$, and the code rate $R$ (in bits per channel use), (9) and (10) enable one to calculate the value of $\left(\frac{E_b}{N_0}\right)_2$ for the second channel, referring to the capacity limitation. To this end, one needs to set $C$ in the LHS of (9) to the code rate $R$, and find the resulting value of $\left(\frac{E_b}{N_0}\right)_2$ which corresponds to the capacity limit. The boundary of the attainable channel region referring to the capacity limit is represented by the continuous curve in Fig. 2 for $R = \frac{1}{3}$ bits per channel use; it is compared to the dashed curve in this figure which represents the boundary of the attainable channel region referring to the cutoff rate limit (see Eq. (8)).

## 2.3 Distance Properties of Ensembles of Turbo-Like Codes

In this paper, we exemplify our numerical results by considering several ensembles of binary linear codes. Due to the outstanding performance of turbo-like codes, our focus is mainly on such ensembles, where we also consider as a reference the ensemble of fully random block codes which achieves capacity under ML decoding. The other ensembles considered in this paper include turbo codes, repeat-accumulate codes and some recent variations.

Bounds on the ML decoding error probability are often based on the distance properties of the considered codes or ensembles (see, e.g., [29] and references therein). The distance spectra and their asymptotic growth rates for various turbo-like ensembles were studied in the literature, e.g., for ensembles of uniformly interleaved repeat-accumulate codes and variations [1, 7, 16], ensembles of uniformly interleaved turbo codes [3, 4, 23, 30], and ensembles of regular and irregular LDPC codes [5, 6, 12, 22]. In this subsection, we briefly present the distance properties of some turbo-like ensembles considered in this paper.

Let us denote by $[\mathcal{C}(n)]$ an ensemble of codes of length $n$. We will also consider a *sequence of ensembles* $[\mathcal{C}(n_1)], [\mathcal{C}(n_2)], \ldots$ where all these ensembles possess a common rate $R$. For a given $(n,k)$ linear code $\mathcal{C}$, let $A_h^{\mathcal{C}}$ (or simply $A_h$) denote the distance spectrum, i.e., the number of codewords of Hamming weight $h$. For a set of codes $[\mathcal{C}(n)]$, we define the *average distance spectrum* as

$$A_h^{[\mathcal{C}(n)]} \triangleq \frac{1}{|[\mathcal{C}(n)]|} \sum_{\mathcal{C} \in [\mathcal{C}(n)]} A_h^{\mathcal{C}}. \tag{12}$$

Let $\Psi_n \triangleq \{\delta : \delta = \frac{h}{n} \text{ for } h = 1, \ldots, n\} = \{\frac{1}{n}, \frac{2}{n}, \ldots, 1\}$ denote the set of *normalized distances*, then the *normalized exponent of the distance spectrum w.r.t. the block length* is defined as

$$r^{\mathcal{C}}(\delta) \triangleq \frac{\ln A_h^{\mathcal{C}}}{n} \quad , \quad r^{[\mathcal{C}(n)]}(\delta) \triangleq \frac{\ln A_h^{[\mathcal{C}(n)]}}{n}. \tag{13}$$

The motivation for this definition lies in the interest to consider the asymptotic case where $n \to \infty$. In this case we define the *asymptotic exponent of the distance spectrum* as

$$r^{[\mathcal{C}]}(\delta) = \lim_{n \to \infty} r^{[\mathcal{C}(n)]}(\delta) . \qquad (14)$$

The input-output weight enumerator (IOWE) of a linear block code is given by a sequence $\{A_{w,h}\}$ designating the number of codewords of Hamming weight $h$ which are encoded by information bits whose Hamming weight is $w$. Referring to ensembles, one considers the average IOWE and distance spectrum over the ensemble. The distance spectrum and the IOWE of linear block codes are useful for the analysis of the block and bit error probabilities, respectively, under ML decoding.

As a reference to all ensembles, we will consider the ensemble of fully random block codes which is capacity-achieving under ML decoding (or 'typical pairs') decoding.

*The ensemble of fully random binary linear block codes*: Consider the ensemble of binary random codes $[\mathcal{RB}]$, where the set $[\mathcal{RB}(n, R)]$ consists of all binary codes of length $n$ and rate $R$. For this ensemble, the following well-known equalities hold:

$$\begin{aligned} A_h^{[\mathcal{RB}(n,R)]} &= \binom{n}{h} 2^{-n(1-R)} \\ r^{[\mathcal{RB}(n,R)]}(\delta) &= \frac{\ln \binom{n}{h}}{n} - (1-R) \ln 2 \\ r^{[\mathcal{RB}(R)]}(\delta) &= H(\delta) - (1-R) \ln 2 \end{aligned} \qquad (15)$$

where $H(x) = -x \ln(x) - (1-x) \ln(1-x)$ designates the binary entropy function to the natural base.

*Non-systematic repeat-accumulate codes*: The ensemble of uniformly interleaved and non-systematic repeat-accumulate (NSRA) codes [7] is defined as follows. The information block of length $N$ is repeated $q$ times by the encoder. The bits are then uniformly reordered by an interleaver of size $qN$, and, finally, encoded by a rate-1 differential encoder (accumulator), i.e., a truncated rate-1 recursive convolutional encoder with a transfer function $1/(1+D)$. The ensemble $[\text{NSRA}(N, q)]$ is defined to be the set of $\frac{(qN)!}{(q!)^N N!}$ RA different codes when considering the different possible permutations of the interleaver.[1] The (average) IOWE of the ensemble of uniformly interleaved RA codes $\mathcal{RA}_q(N)$ was originally derived in [7, Section 5], and it is given by

$$A_{w,h}^{\text{NSRA}(N,q)} = \frac{\binom{N}{w} \binom{qN-h}{\lfloor \frac{qw}{2} \rfloor} \binom{h-1}{\lceil \frac{qw}{2} \rceil - 1}}{\binom{qN}{qw}} \qquad (16)$$

and therefore the distance spectrum of the ensemble is given by

$$A_h^{\text{NSRA}(N,q)} = \sum_{w=1}^{\min(N, \lfloor \frac{2h}{q} \rfloor)} \frac{\binom{N}{w} \binom{qN-h}{\lfloor \frac{qw}{2} \rfloor} \binom{h-1}{\lceil \frac{qw}{2} \rceil - 1}}{\binom{qN}{qw}}, \quad \left\lceil \frac{q}{2} \right\rceil \leq h \leq qN - \left\lfloor \frac{q}{2} \right\rfloor$$

$A_h^{\text{NSRA}(N,q)} = 0$ for $1 \leq h < \lceil \frac{q}{2} \rceil$, and $A_0^{\text{NSRA}(N,q)} = 1$ since the all-zero vector is always a codeword of a linear code. The asymptotic exponent of the distance spectrum of this ensemble is given by

---

[1] There are $(qN)!$ ways to place $qN$ bits. However, permuting the $q$ repetitions of any of the $N$ information bits does not affect the result of the interleaving, so there are $\frac{(qN)!}{(q!)^N}$ possible ways for the interleaving. Strictly speaking, by permuting the $N$ information bits, the vector space of the code does not change, which then yields that there are $\frac{(qN)!}{(q!)^N N!}$ distinct RA codes of dimension $k$ and number of repetitions $q$.

(see [17])

$$
\begin{aligned}
r^{[\text{NSRA}(q)]}(\delta) &\triangleq \lim_{N \to \infty} r^{[\text{NSRA}(N,q)]}(\delta) \\
&= \max_{0 \leq u \leq \min(2\delta, 2-2\delta)} \left\{ -\left(1 - \frac{1}{q}\right) H(u) + (1-\delta) H\left(\frac{u}{2(1-\delta)}\right) + \delta H\left(\frac{u}{2\delta}\right) \right\}. \quad (17)
\end{aligned}
$$

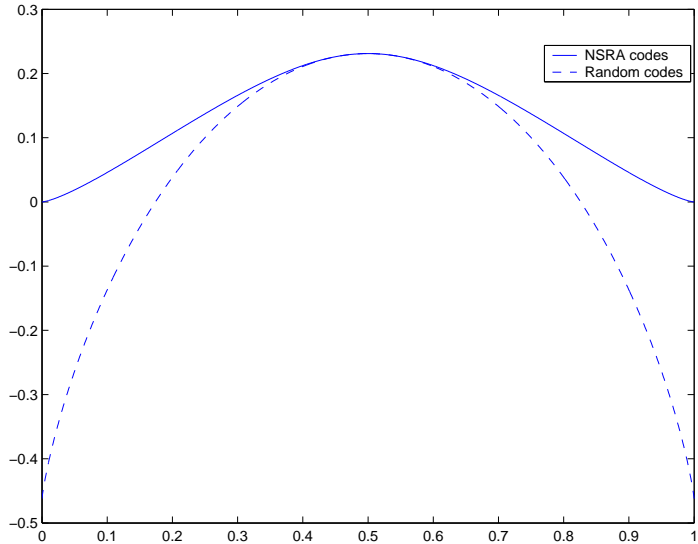

Figure 3: Plot of the normalized asymptotic distance spectra for the ensembles of fully random block codes and uniformly interleaved and non-systematic repeat-accumulate (NSRA) codes of rate $\frac{1}{3}$ bits per channel use. The curves are depicted as a function of the normalized Hamming weight ($\delta$), and their calculations are based on (15) and (17).

The IOWEs and distance spectra of various ensembles of irregular repeat-accumulate (IRA) and accumulate-repeat-accumulate (ARA) codes are derived in [1, 16].

## 2.4 The DS2 Bound for a Single MBIOS Channel

The bounding technique of Duman and Salehi [9, 10] originates from the 1965 Gallager bound [13] which states that the conditional ML decoding error probability $P_{e|m}$ given that a codeword $\underline{x}^m$ (of block length $n$) is transmitted is upper-bounded by

$$
P_{e|m} \leq \sum_{\underline{y}} p_n\left(\underline{y} | \underline{x}^m\right) \left( \sum_{m' \neq m} \left( \frac{p_n(\underline{y}|\underline{x}^{m'})}{p_n(\underline{y}|\underline{x}^m)} \right)^\lambda \right)^\rho \quad \lambda, \rho \geq 0 \quad (18)
$$

where $p_n(\underline{y}|\underline{x})$ designates the conditional *pdf* of the communication channel to obtain an $n$-length sequence $\underline{y}$ at the channel output, given the $n$-length input sequence $\underline{x}$.

Unfortunately, this upper bound is not calculable in terms of the distance spectrum of the code ensemble, except for the particular cases of ensembles of fully random block codes and orthogonal codes transmitted over a memoryless channel, and the special case where $\rho = 1, \lambda = 0.5$ in which the bound reduces to the union-Bhattacharyya bound. With the intention of alleviating the difficulty of calculating the bound for specific codes and ensembles, we introduce the function $\Psi_n^{(m)}(\underline{y})$ which is an arbitrary probability tilting measure. This function may depend in general on the index $m$ of the transmitted codeword [32], and is a non-negative function which satisfies the equality $\int_{\underline{y}} \Psi_n^{(m)}(\underline{y}) \, d\underline{y} = 1$. The upper bound in (18) can be rewritten in the following equivalent form:

$$P_{\mathrm{e}|m} \leq \sum_{\underline{y}} \Psi_n^{(m)}(\underline{y}) \left( \Psi_n^{(m)}(\underline{y})^{-\frac{1}{\rho}} p_n\left(\underline{y}|\underline{x}^m\right) \sum_{m' \neq m} \left( \frac{p_n(\underline{y}|\underline{x}^{m'})}{p_n(\underline{y}|\underline{x}^m)} \right)^{\lambda} \right)^{\rho} \quad \lambda, \rho \geq 0. \quad (19)$$

Recalling that $\Psi_n^{(m)}$ is a probability measure, we invoke Jensen's inequality in (19) which gives

$$P_{\mathrm{e}|m} \leq \left( \sum_{m' \neq m} \sum_{\underline{y}} \Psi_n^{(m)}(\underline{y})^{1-\frac{1}{\rho}} p_n(\underline{y}|\underline{x}^m)^{\frac{1}{\rho}} \left( \frac{p_n(\underline{y}|\underline{x}^{m'})}{p_n(\underline{y}|\underline{x}^m)} \right)^{\lambda} \right)^{\rho}, \quad \begin{array}{c} 0 \leq \rho \leq 1 \\ \lambda \geq 0 \end{array} \quad (20)$$

which is the DS2 bound. This expression can be simplified (see, e.g., [32]) for the case of a single memoryless channel where

$$p_n(\underline{y}|\underline{x}) = \prod_{i=1}^{n} p(y_i|x_i).$$

Let us consider probability tilting measures $\Psi_n^{(m)}(\underline{y})$ which can be factorized into the form

$$\Psi_n^{(m)}(\underline{y}) = \prod_{i=1}^{n} \psi^{(m)}(y_i)$$

recalling that the function $\psi^{(m)}$ may depend on the transmitted codeword $\underline{x}^m$. In this case, the bound in (20) is calculable in terms of the distance spectrum of the code, thus not requiring the fine details of the code structure.

Let $\mathcal{C}$ be a binary linear block code whose block length is $n$, and let its distance spectrum be given by $\{A_h\}_{h=0}^n$. Consider the case where the transmission takes place over an MBIOS channel. By partitioning the code into subcodes of constant Hamming weight, let $\mathcal{C}_h$ be the set which includes all the codewords of $\mathcal{C}$ with Hamming weight $h$ and the all-zero codeword. Note that this forms a partitioning of a linear code into subcodes which are in general non-linear. We apply the DS2 bound on the conditional ML decoding error probability (given the all-zero codeword is transmitted), and finally use the union bound w.r.t. the subcodes $\{\mathcal{C}_h\}$ in order to obtain an upper bound on the ML decoding error probability of the code $\mathcal{C}$. Referring to the constant Hamming weight subcode $\mathcal{C}_h$, the bound (20) gives

$$P_{\mathrm{e}|0}(h) \leq (A_h)^{\rho} \left\{ \left( \sum_y \psi(y)^{1-\frac{1}{\rho}} p(y|0)^{\frac{1}{\rho}} \right)^{n-h} \left( \sum_y \psi(y)^{1-\frac{1}{\rho}} p(y|0)^{\frac{1-\lambda\rho}{\rho}} p(y|1)^{\lambda} \right)^{h} \right\}^{\rho} \quad \begin{array}{c} 0 \leq \rho \leq 1 \\ \lambda \geq 0 \end{array} . (21)$$

Clearly, for an MBIOS channel with continuous output, the sums in (21) are replaced by integrals. In order to obtain the tightest bound within this form, the probability tilting measure $\psi$ and the parameters $\lambda$ and $\rho$ are optimized. The optimization of $\psi$ is based on calculus of variations, and is independent of the distance spectrum.

Due to the symmetry of the channel and the linearity of the code $\mathcal{C}$, the decoding error probability of $\mathcal{C}$ is independent of the transmitted codeword. Since the code $\mathcal{C}$ is the union of the subcodes $\{\mathcal{C}_h\}$, the union bound provides an upper bound on the ML decoding error probability of $\mathcal{C}$ which is expressed as the sum of the conditional decoding error probabilities of the subcodes $\mathcal{C}_h$ given that the all-zero codeword is transmitted. Let $d_{\min}$ be the minimum distance of the code $\mathcal{C}$, and $R$ be the rate of the code $\mathcal{C}$. Based on the linearity of the code, the geometry of the Voronoi regions implies that one can ignore those subcodes whose Hamming weights are above $n(1-R)$ (see [2]). Hence, the expurgated union bound gets the form

$$P_{\text{e}} \leq \sum_{h=d_{\min}}^{n(1-R)} P_{\text{e}|0}(h). \tag{22}$$

For the bit error probability, one may partition a binary linear block code $\mathcal{C}$ into subcodes w.r.t. the Hamming weights of the information bits and the coded bits. Let $\mathcal{C}_{w,h}$ designate the subcode which includes the all-zero codeword and all the codeowrds of $\mathcal{C}$ whose Hamming weight is $h$ and their information bits have Hamming weight $w$. An upper bound on the bit error probability of the code $\mathcal{C}$ is performed by calculating the DS2 upper bound on the conditional bit error probability for each subcode $\mathcal{C}_{w,h}$ (given that the all-zero codeword is transmitted), and applying the union bound over all these subcodes. Note that the number of these subcodes is at most quadratic in the block length of the code, so taking the union bound w.r.t. these subcodes does not affect the asymptotic tightness of the overall bound. Let $\{A_{w,h}\}$ designate the IOWE of the code $\mathcal{C}$ whose block length and dimension are equal to $n$ and $k$, respectively. The conditional DS2 bound on the bit error probability was demonstrated in [28, 29] to be identical to the DS2 bound on the block error probability, except that the distance spectrum of the code

$$A_h = \sum_{w=0}^{k} A_{w,h}, \quad h = 0, \ldots, n \tag{23}$$

appearing in the RHS of (21) is replaced by

$$A'_h \triangleq \sum_{w=0}^{k} \left(\frac{w}{k}\right) A_{w,h}, \quad h = 0, \ldots, n. \tag{24}$$

Since $A'_h \leq A_h$ then, as expected, the upper bound on the bit error probability is smaller than the upper bound on the block error probability.

Finally, note that the DS2 bound is also applicable to ensembles of linear codes. To this end, one simply needs to replace the distance spectrum or the IOWE of a code by the average quantities over this ensemble. This follows easily by invoking Jensen's inequality to the RHS of (21) which yields that $\mathbb{E}[(A_h)^\rho] \leq (\mathbb{E}[A_h])^\rho$ for $0 \leq \rho \leq 1$.

The DS2 bound for a single channel is discussed in further details in [9, 28, 32] and the tutorial paper [29, Chapter 4].

## 3 The Generalized DS2 Bound for Independent Parallel Channels

In this section, we generalize the DS2 bound to independent parallel channels, and optimize the related probability tilting measures which enable to obtain the tightest bound within this form.

### 3.1 Derivation of the New Bound

Let us assume that the communication takes place over $J$ statistically independent parallel channels where each one of the individual channels is a memoryless binary-input output-symmetric (MBIOS) with antipodal signaling, i.e., $p(y|x=1) = p(-y|x=-1)$. We start our discussion by considering the case of a specific channel assignment. By assuming that all $J$ channels are independent and MBIOS, we may factor the transition probability as

$$p_n\left(\underline{y}|\underline{x}^m\right) = \prod_{j=1}^{J} \prod_{i \in \mathcal{I}(j)} p(y_i | x_i^{(m)}; j) \tag{25}$$

which we can plug into (20) to get a DS2 bound suitable for the case of parallel channels. In order to get a bound which depends on one-dimensional sums (or one-dimensional integrals), we impose a restriction on the tilting measure $\Psi_n^{(m)}(\cdot)$ in (20) so that it can be expressed as a $J$-fold product of one-dimensional probability tilting measures, i.e.,

$$\Psi_n^{(m)}(\underline{y}) = \prod_{j=1}^{J} \prod_{i \in \mathcal{I}(j)} \psi^{(m)}(y_i; j). \tag{26}$$

Considering a binary linear block code $\mathcal{C}$, the conditional decoding error probability does not depend on the transmitted codeword, so $P_\mathrm{e} \triangleq \frac{1}{M} \sum_{m=0}^{M-1} P_{\mathrm{e}|m} = P_{\mathrm{e}|0}$ where w.o.l.o.g., one can assume that the all-zero vector is the transmitted codeword.

The channel mapper for the $J$ independent parallel channels is assumed to transmit the bits whose indices are included in the subset $\mathcal{I}(j)$ over the $j$-th channel where the subsets $\{\mathcal{I}(j)\}$ constitute a disjoint partitioning of the set of indices $\{1, 2, \ldots, n\}$.

Following the notation in [21], let $A_{h_1, h_2, \ldots, h_J}$ designate the *split weight enumerator* of the binary linear block code, defined as the number of codewords of Hamming weight $h_j$ within the $J$ disjoint subsets $\mathcal{I}(j)$ for $j = 1 \ldots J$. By substituting (25) and (26) in (20), we obtain

$$
\begin{aligned}
P_\mathrm{e} &= P_{\mathrm{e}|0} \\
&\leq \left\{ \sum_{h_1=0}^{|\mathcal{I}(1)|} \cdots \sum_{h_J=0}^{|\mathcal{I}(J)|} \sum_{\underline{y}} A_{h_1,h_2,\ldots,h_J} \prod_{j=1}^{J} \prod_{i \in \mathcal{I}(j)} \psi(y_i;j)^{1-\frac{1}{\rho}} p(y_i|0;j)^{\frac{1}{\rho}} \left( \frac{p(y_i|x_i;j)}{p(y_i|0;j)} \right)^\lambda \right\}^\rho \\
&= \left\{ \sum_{h_1=0}^{|\mathcal{I}(1)|} \cdots \sum_{h_J=0}^{|\mathcal{I}(J)|} A_{h_1,h_2,\ldots,h_J} \prod_{j=1}^{J} \prod_{i \in \mathcal{I}(j)} \sum_{y_i} \psi(y_i;j)^{1-\frac{1}{\rho}} p(y_i|0;j)^{\frac{1}{\rho}} \left( \frac{p(y_i|x_i;j)}{p(y_i|0;j)} \right)^\lambda \right\}^\rho \\
&= \left\{ \sum_{h_1=0}^{|\mathcal{I}(1)|} \cdots \sum_{h_J=0}^{|\mathcal{I}(J)|} A_{h_1,h_2,\ldots,h_J} \prod_{j=1}^{J} \left( \sum_{y} \psi(y;j)^{1-\frac{1}{\rho}} p(y|0;j)^{\frac{1-\lambda\rho}{\rho}} p(y|1;j)^\lambda \right)^{h_j} \right. \\
&\qquad \left. \prod_{j=1}^{J} \left( \sum_{y} \psi(y;j)^{1-\frac{1}{\rho}} p(y|0;j)^{\frac{1}{\rho}} \right)^{|\mathcal{I}(j)|-h_j} \right\}^\rho, \quad \begin{array}{c} 0 \leq \rho \leq 1 \\ \lambda \geq 0 \end{array}. \tag{27}
\end{aligned}
$$

We note that the bound in (27) is valid for a specific assignment of bits to the parallel channels. For structured codes or ensembles, the split weight enumerator is in general not available when considering specific assignments. As a result of this, we continue the derivation of the bound by using the random assignment approach. Let us designate $n_j \triangleq |\mathcal{I}(j)|$ to be the cardinality of the set $\mathcal{I}(j)$, so $E[n_j] = \alpha_j n$ is the expected number of bits assigned to channel no. $j$ (where $j = 1, 2, \ldots, J$). Averaging (27) with respect to all possible channel assignments, we get the following bound on the average ML decoding error probability:

$$P_e \leq \mathbf{E}\left\{\sum_{h_1=0}^{n_1} \cdots \sum_{h_J=0}^{n_J} A_{h_1,h_2,\ldots,h_J} \prod_{j=1}^{J}\left(\sum_y \psi(y;j)^{1-\frac{1}{\rho}} p(y|0;j)^{\frac{1-\lambda\rho}{\rho}} p(y|1;j)^{\lambda}\right)^{h_j}\right.$$

$$\left.\prod_{j=1}^{J}\left(\sum_y \psi(y;j)^{1-\frac{1}{\rho}} p(y|0;j)^{\frac{1}{\rho}}\right)^{n_j-h_j}\right\}^{\rho}$$

$$= \sum_{\substack{n_j \geq 0 \\ \sum_j n_j = n}} \left\{\sum_{h_1=0}^{n_1} \cdots \sum_{h_J=0}^{n_J} A_{h_1,h_2,\ldots,h_J} \prod_{j=1}^{J}\left(\sum_y \psi(y;j)^{1-\frac{1}{\rho}} p(y|0;j)^{\frac{1-\lambda\rho}{\rho}} p(y|1;j)^{\lambda}\right)^{h_j}\right.$$

$$\left.\prod_{j=1}^{J}\left(\sum_y \psi(y;j)^{1-\frac{1}{\rho}} p(y|0;j)^{\frac{1}{\rho}}\right)^{n_j-h_j}\right\}^{\rho} P_{\underline{N}}(\underline{n}) \qquad (28)$$

where $P_{\underline{N}}(\underline{n})$ designates the probability mass function of the discrete random vector $\underline{N} \triangleq (n_1, \ldots, n_J)$. After applying Jensen's inequality to the RHS of (28) and changing the order of summation, we get

$$P_e \leq \left\{\sum_{\substack{n_j \geq 0 \\ \sum n_j = n}} \sum_{h=0}^{n} \sum_{\substack{h_1 \leq n_1, \ldots, h_J \leq n_J \\ h_1 + \ldots + h_J = h}} A_{h_1,h_2,\ldots,h_J} P_{\underline{N}}(\underline{n})\right.$$

$$\prod_{j=1}^{J}\left(\sum_y \psi(y;j)^{1-\frac{1}{\rho}} p(y|0;j)^{\frac{1-\lambda\rho}{\rho}} p(y|1;j)^{\lambda}\right)^{h_j}$$

$$\left.\prod_{j=1}^{J}\left(\sum_y \psi(y;j)^{1-\frac{1}{\rho}} p(y|0;j)^{\frac{1}{\rho}}\right)^{n_j-h_j}\right\}^{\rho}, \quad \begin{matrix}0 \leq \rho \leq 1 \\ \lambda \geq 0\end{matrix} . \qquad (29)$$

Let the vector $\underline{H} = (h_1, \ldots, h_J)$ be the vector of partial Hamming weights referring to the bits transmitted over each channel ($n_j$ bits are transmitted over channel no. $j$, so $0 \leq h_j \leq n_j$). Clearly, $\sum_{j=1}^{J} h_j = h$ is the overall Hamming weight of a codeword in $\mathcal{C}$. Due to the random assignment of the code bits to the parallel channels, we get

$$P_{\underline{N}}(\underline{n}) = \binom{n}{n_1, n_2, \ldots, n_J} \alpha_1^{n_1} \alpha_2^{n_2} \ldots \alpha_J^{n_J}$$

$$P_{\underline{H}|\underline{N}}(\underline{h}|\underline{n}) = \frac{\binom{h}{h_1,\ldots,h_J}\binom{n-h}{n_1-h_1,\ldots,n_J-h_J}}{\binom{n}{n_1,\ldots,n_J}}$$

$$A_{h_1,h_2,\ldots,h_J} P_{\underline{N}}(\underline{n})$$
$$= A_h\, P_{\underline{H}|\underline{N}}(\underline{h}|\underline{n})\, P_{\underline{N}}(\underline{n})$$
$$= A_h\, \alpha_1^{n_1} \alpha_2^{n_2} \ldots \alpha_J^{n_J} \binom{h}{h_1, \ldots, h_J}\binom{n-h}{n_1-h_1, \ldots, n_J-h_J} \qquad (30)$$

and the substitution of (30) in (29) gives

$$
P_{\text{e}} \leq \left\{ \sum_{\substack{n_j \geq 0 \\ \sum n_j = n}} \sum_{h=0}^{n} A_h \sum_{\substack{h_1 \leq n_1, \ldots, h_J \leq n_J \\ h_1 + \ldots + h_J = h}} \binom{h}{h_1, h_2, \ldots, h_J} \right.
$$
$$
\binom{n-h}{n_1 - h_1, n_2 - h_2, \ldots, n_J - h_J}
$$
$$
\prod_{j=1}^{J} \left( \alpha_j \sum_y \psi(y; j)^{1-\frac{1}{\rho}} p(y|0;j)^{\frac{1-\lambda\rho}{\rho}} p(y|1;j)^{\lambda} \right)^{h_j}
$$
$$
\left. \prod_{j=1}^{J} \left( \alpha_j \sum_y \psi(y; j)^{1-\frac{1}{\rho}} p(y|0;j)^{\frac{1}{\rho}} \right)^{n_j - h_j} \right\}^{\rho}.
$$

Let $k_j \triangleq n_j - h_j$ for $j = 1, 2, \ldots, J$, then by changing the order of summation in the above bound, we obtain

$$
P_{\text{e}} \leq \left\{ \sum_{h=0}^{n} A_h \sum_{\substack{h_1, \ldots, h_J \geq 0 \\ h_1 + \ldots + h_J = h}} \binom{h}{h_1, h_2, \ldots, h_J} \prod_{j=1}^{J} \left( \alpha_j \sum_y \psi(y; j)^{1-\frac{1}{\rho}} p(y|0;j)^{\frac{1-\lambda\rho}{\rho}} p(y|1;j)^{\lambda} \right)^{h_j} \right.
$$
$$
\left. \sum_{\substack{k_1, \ldots, k_J \geq 0 \\ k_1 + \ldots + k_J = n-h}} \binom{n-h}{k_1, k_2, \ldots, k_J} \prod_{j=1}^{J} \left( \alpha_j \sum_y \psi(y; j)^{1-\frac{1}{\rho}} p(y|0;j)^{\frac{1}{\rho}} \right)^{k_j} \right\}^{\rho}
$$

Since $\sum_{j=1}^{J} h_j = h$ and $\sum_{j=1}^{J} k_j = n - h$, the use of the multinomial formula gives

$$
P_{\text{e}} \leq \left\{ \sum_{h=0}^{n} A_h \left( \sum_{j=1}^{J} \alpha_j \sum_y \psi(y; j)^{1-\frac{1}{\rho}} p(y|0;j)^{\frac{1-\lambda\rho}{\rho}} p(y|1;j)^{\lambda} \right)^{h} \right.
$$
$$
\left. \left( \sum_{j=1}^{J} \alpha_j \sum_y \psi(y; j)^{1-\frac{1}{\rho}} p(y|0;j)^{\frac{1}{\rho}} \right)^{n-h} \right\}^{\rho} \quad \begin{array}{c} 0 \leq \rho \leq 1 \\ \lambda \geq 0 \\ \sum_y \psi(y;j) = 1 \\ j = 1 \ldots J \end{array} \quad (31)
$$

which forms a generalization of the DS2 bound for independent parallel channels, where the bound is averaged over all possible channel assignments. This result can be applied to specific codes as well as to structured ensembles for which the average distance spectrum $\overline{A_h}$ is known. In this case, the average ML decoding error probability $\overline{P_{\text{e}}}$ is obtained by replacing $A_h$ in (31) with $\overline{A_h}$.[2]

In the continuation of this section, we propose an equivalent version of the generalized DS2 bound for parallel channels where this equivalence follows the lines in [29, 32]. Rather than relying on a probability (i.e., normalized) tilting measure, the bound will be expressed in terms of an unnormalized tilting measure which is an arbitrary non-negative function. This version will be helpful later for the discussion on the connection between the DS2 bound and the 1961 Gallager bound for parallel channels, and also for the derivation of some particular cases of the DS2 bound. We begin by expressing the DS2 bound using the un-normalized tilting measure $G_n^{(m)}$ which is related to $\Psi_n^{(m)}$ by

$$\Psi_n^{(m)}(\underline{y}) = \frac{G_n^{(m)}(\underline{y})p_n(\underline{y}|\underline{x}^m)}{\sum_{\underline{y}'} G_n^{(m)}(\underline{y}')p_n(\underline{y}'|\underline{x}^m)} \ . \tag{32}$$

---

[2]This can be shown by noting that the function $f(t) = t^\rho$ is convex for $0 \leq \rho \leq 1$ and by invoking Jensen's inequality in (31).

Substituting (32) in (20) gives

$$P_{\mathrm{e}|m} \leq \left( \sum_{\underline{y}} G_n^{(m)}(\underline{y})p_n(\underline{y}|\underline{x}^m) \right)^{1-\rho} \left\{ \sum_{m' \neq m} \sum_{\underline{y}} G_n^{(m)}(\underline{y})^{1-\frac{1}{\rho}} p_n(\underline{y}|\underline{x}^m) \left( \frac{p_n(\underline{y}|\underline{x}^{m'})}{p_n(\underline{y}|\underline{x}^m)} \right)^{\lambda} \right\}^{\rho}, \quad \begin{array}{c} 0 \leq \rho \leq 1 \\ \lambda \geq 0 \end{array}.$$

As before, we assume that $G_n^{(m)}$ can be factored in the product form

$$G_n^{(m)}(\underline{y}) = \prod_{j=1}^{J} \prod_{i \in \mathcal{I}(j)} g(y_i; j).$$

Following the algebraic steps in (27)-(31) and averaging as before also over all the codebooks of the ensemble, we obtain the following upper bound on the ML decoding error probability:

$$\overline{P_{\mathrm{e}}} = \overline{P_{\mathrm{e}|0}} \leq \left\{ \sum_{h=0}^{n} \overline{A}_h \left[ \sum_{j=1}^{J} \alpha_j \left( \sum_{y} g(y;j)^{1-\frac{1}{\rho}} p(y|0;j)^{1-\lambda} p(y|1;j)^{\lambda} \right) \right. \right.$$
$$\left( \sum_{y} g(y;j)p(y|0;j) \right)^{\frac{1-\rho}{\rho}} \right]^{h} \left[ \sum_{j=1}^{J} \alpha_j \left( \sum_{y} g(y;j)^{1-\frac{1}{\rho}} p(y|0;j) \right) \right.$$
$$\left. \left. \left( \sum_{y} g(y;j)p(y|0;j) \right)^{\frac{1-\rho}{\rho}} \right]^{n-h} \right\}^{\rho}, \quad \begin{array}{c} 0 \leq \rho \leq 1 \\ \lambda \geq 0 \end{array}. \tag{33}$$

Note that the generalized DS2 bound as derived in this subsection is applied to the whole code (i.e., the optimization of the tilting measures refers to the whole code and is performed only once for each of the $J$ channels). In the next subsection, we consider the partitioning of the code to constant Hamming weight subcodes, and then apply the union bound. For every such subcode, we rely on the conditional DS2 bound (given the all-zero codeword is transmitted), and optimize the $J$ tilting measures *separately*. The total number of subcodes does not exceed the block length of the code (or ensemble), and hence the use of the union bound in this case does not degrade the related error exponent of the overall bound, but on the other hand, the optimized tilting measures are tailored for each of the constant-Hamming weight subcodes, a process which can only improve the exponential behavior of the resulting bound.

### 3.2 Optimization of the Tilting Measures for the Generalized DS2 Bound

In this section, we find optimized tilting measures $\{\psi(\cdot; j)\}_{j=1}^{J}$ which minimize the DS2 bound (31). The following calculation generalizes the analysis in [32] for a single channel to the considered case of an arbitrary number ($J$) of independent parallel MBIOS channels.

Let $\mathcal{C}$ be a binary linear block code of length $n$. Following the derivation in [21, 32], we partition the code $\mathcal{C}$ to constant Hamming weight subcodes $\{\mathcal{C}_h\}_{h=0}^{n}$, where $\mathcal{C}_h$ includes all the codewords of weight $h$ ($h = 0, \ldots, n$) as well as the all-zero codeword. Let $P_{\text{e}|0}(h)$ denote the conditional block error probability of the subcode $\mathcal{C}_h$ under ML decoding, given that the all-zero codeword is transmitted. Based on the union bound, we get

$$P_{\text{e}} \leq \sum_{h=0}^{n} P_{\text{e}|0}(h). \tag{34}$$

As the code $\mathcal{C}$ is linear, $P_{\text{e}|0}(h) = 0$ for $h = 0, 1, \ldots, d_{\min} - 1$ where $d_{\min}$ denotes the minimum distance of the code $\mathcal{C}$. The generalization of the DS2 bound in (31) gives the following upper bound on the conditional error probability of the subcode $\mathcal{C}_h$:

$$P_{\text{e}|0}(h) \leq (A_h)^\rho \left\{ \left( \sum_{j=1}^{J} \alpha_j \sum_y \psi(y;j)^{1-\frac{1}{\rho}} p(y|0;j)^{\frac{1-\lambda\rho}{\rho}} p(y|1;j)^\lambda \right)^\delta \right.$$

$$\left. \left( \sum_{j=1}^{J} \alpha_j \sum_y \psi(y;j)^{1-\frac{1}{\rho}} p(y|0;j)^{\frac{1}{\rho}} \right)^{1-\delta} \right\}^{n\rho}, \quad 0 \leq \rho \leq 1, \quad \lambda \geq 0, \quad \delta \triangleq \frac{h}{n}. \tag{35}$$

Note that in this case, the set of probability tilting measures $\{\psi(\cdot;j)\}_{j=1}^{J}$ may also depend on the Hamming weight ($h$) of the subcode (or equivalently on $\delta$). This is the result of performing the optimization on every individual constant-Hamming subcode instead of the whole code.

This generalization of the DS2 bound can be written equivalently in the exponential form

$$P_{\text{e}|0}(h) \leq e^{-n E_\delta^{\text{DS2}}(\lambda, \rho, J, \{\alpha_j\})}, \quad 0 \leq \rho \leq 1, \quad \lambda \geq 0, \quad \delta \triangleq \frac{h}{n}. \tag{36}$$

where

$$E_\delta^{\text{DS2}}(\lambda, \rho, J, \{\alpha_j\}) \triangleq -\rho r^\mathcal{C}(\delta) - \rho \delta \ln \left( \sum_{j=1}^{J} \alpha_j \sum_y \psi(y;j)^{1-\frac{1}{\rho}} p(y|0;j)^{\frac{1-\lambda\rho}{\rho}} p(y|1;j)^\lambda \right)$$

$$- \rho(1-\delta) \ln \left( \sum_{j=1}^{J} \alpha_j \sum_y \psi(y;j)^{1-\frac{1}{\rho}} p(y|0;j)^{\frac{1}{\rho}} \right) \tag{37}$$

and $r^\mathcal{C}(\delta)$ designates the normalized exponent of the distance spectrum as in (14).

Let

$$g_1(y;j) \triangleq p(y|0;j)^{\frac{1}{\rho}}, \quad g_2(y;j) \triangleq p(y|0;j)^{\frac{1}{\rho}} \left( \frac{p(y|1;j)}{p(y|0;j)} \right)^\lambda \tag{38}$$

then, for a given pair of $\lambda$ and $\rho$ (where $\lambda \geq 0$ and $0 \leq \rho \leq 1$), we need to minimize

$$\delta \ln \left( \sum_{j=1}^{J} \alpha_j \sum_y \psi(y;j)^{1-\frac{1}{\rho}} g_2(y;j) \right) + (1-\delta) \ln \left( \sum_{j=1}^{J} \alpha_j \sum_y \psi(y;j)^{1-\frac{1}{\rho}} g_1(y;j) \right)$$

over the set of non-negative functions $\psi(\,\cdot\,;j)$ satisfying the constraints

$$\sum_y \psi(y;j) = 1, \quad j = 1 \ldots J. \tag{39}$$

To this end, calculus of variations provides the following set of equations:

$$\psi(y;j)^{-\frac{1}{\rho}} \left( \frac{\alpha_j(1-\delta)(1-\frac{1}{\rho})g_1(y;j)}{\sum_y \sum_{j=1}^J \alpha_j \psi(y;j)^{1-\frac{1}{\rho}} g_1(y;j)} \right.$$
$$\left. + \frac{\alpha_j \delta(1-\frac{1}{\rho})g_2(y;j)}{\sum_y \sum_{j=1}^J \alpha_j \psi(y;j)^{1-\frac{1}{\rho}} g_2(y;j)} \right) + \xi_j = 0, \quad j=1,\ldots,J \quad (40)$$

where $\xi_j$ is a Lagrange multiplier. The solution of (40) is given in the following implicit form:

$$\psi(y;j) = \bigl(k_{1,j} g_1(y;j) + k_{2,j} g_2(y;j)\bigr)^\rho, \qquad k_{1,j}, k_{2,j} \geq 0, \quad j=1,\ldots,J$$

where

$$\frac{k_{2,j}}{k_{1,j}} = \frac{\delta}{1-\delta} \frac{\sum_{j=1}^J \sum_{y\in\mathcal{Y}} \alpha_j \psi(y;j)^{1-\frac{1}{\rho}} g_1(y;j)}{\sum_{j=1}^J \sum_{y\in\mathcal{Y}} \alpha_j \psi(y;j)^{1-\frac{1}{\rho}} g_2(y;j)} . \quad (41)$$

We note that $k \triangleq \frac{k_{2,j}}{k_{1,j}}$ in the RHS of (41) is *independent* of $j$. Thus, the substitution $\beta_j \triangleq k_{1,j}^\rho$ gives that the optimal tilting measures can be expressed as

$$\psi(y;j) = \beta_j \bigl(g_1(y;j) + k g_2(y;j)\bigr)^\rho$$
$$= \beta_j \, p(y|0;j) \left[1 + k\left(\frac{p(y|1;j)}{p(y|0;j)}\right)^\lambda\right]^\rho \quad y\in\mathcal{Y} \quad j=1,\ldots,J. \quad (42)$$

By plugging (38) into (41) we obtain

$$k = \frac{\delta}{1-\delta} \frac{\sum_{j=1}^J \sum_{y\in\mathcal{Y}} \left\{ \alpha_j \beta_j^{1-\frac{1}{\rho}} p(y|0;j) \left[1 + k\left(\frac{p(y|1;j)}{p(y|0;j)}\right)^\lambda\right]^{\rho-1} \right\}}{\sum_{j=1}^J \sum_{y\in\mathcal{Y}} \left\{ \alpha_j \beta_j^{1-\frac{1}{\rho}} p(y|0;j) \left(\frac{p(y|1;j)}{p(y|0;j)}\right)^\lambda \left[1 + k\left(\frac{p(y|1;j)}{p(y|0;j)}\right)^\lambda\right]^{\rho-1} \right\}} \quad (43)$$

and from (38) and (39), $\beta_j$ which is the appropriate factor normalizing the probability tilting measure $\psi(\cdot;j)$ in (42) is given by

$$\beta_j = \left[\sum_{y\in\mathcal{Y}} p(y|0;j)\left(1 + k\left(\frac{p(y|1;j)}{p(y|0;j)}\right)^\lambda\right)^\rho\right]^{-1}, \quad j=1,\ldots,J. \quad (44)$$

Note that the implicit equation for $k$ in (43) and the normalization coefficients in (44) generalize the results derived in [28, Appendix A] (where $k$ is replaced there by $\alpha$). The key result here is that the values of $\frac{k_{2,j}}{k_{1,j}}$ in (41) are independent of $j$ (where $j \in \{1,2,\ldots,J\}$), a property which significantly simplifies the optimization process of the $J$ tilting measures, and leads to the result in (42).

For the numerical calculation of the bound in (35) as a function of the normalized Hamming weight $\delta \triangleq \frac{h}{n}$, and for a fixed pair of $\lambda$ and $\rho$ (where $\lambda \geq 0$ and $0 \leq \rho \leq 1$), we find the optimized tilting measures in (42) by first assuming an initial vector $\overline{\beta^{(0)}} = (\beta_1,\ldots,\beta_J)$ and then iterating between (43) and (44) until we get a fixed point for these equations. For a fixed $\delta$, we need to optimize numerically the bound in (36) w.r.t. the two parameters $\lambda$ and $\rho$.

### 3.3 Statement of the Main Result Derived in Section 3

The analysis in this section leads to the following theorem:

**Theorem 1 (Generalized DS2 bound for independent parallel MBIOS channels).** Consider the transmission of binary linear block codes (or ensembles) over a set of $J$ independent parallel MBIOS channels. Let the *pdf* of the $j^{\text{th}}$ MBIOS channel be given by $p(\cdot|0;j)$ where due to the symmetry of the binary-input channels $p(y|0;j) = p(-y|1;j)$. Assume that the coded bits are randomly and independently assigned to these channels, where each bit is transmitted over one of the $J$ MBIOS channels. Let $\alpha_j$ be the a-priori probability of transmitting a bit over the $j^{\text{th}}$ channel ($j = 1, 2, \ldots, J$), so that $\alpha_j \geq 0$ and $\sum_{j=1}^{J} \alpha_j = 1$. Then, the generalization of the DS2 bound in (31) provides an upper bound on the ML decoding error probability when the bound is taken over the whole code. By partitioning the code into constant Hamming-weight subcodes, (35) forms an upper bound on the conditional ML decoding error probability for each of these subcodes, given that the all-zero codeword is transmitted, and (34) forms an upper bound on the block error probability of the whole code (or ensemble). For an arbitrary constant Hamming weight subcode, the optimized set of probability tilting measures $\{\psi(\cdot;j)\}_{j=1}^{J}$ which attains the minimal value of the conditional upper bound in (35) is given by the set of equations in (42)–(44).

## 4 Generalization of the 1961 Gallager Bound for Parallel Channels and Its Connection to the Generalized DS2 Bound

The 1961 Gallager bound for a single MBIOS channel was derived in [12], and the generalization of the bound for parallel MBIOS channels was derived by Liu et al. [21]. In the following, we outline its derivation in [21] which serves as a preliminary step towards the discussion of its relation to the generalized DS2 bound from Section 3. In this section, we optimize the probability tilting measures which are related to the 1961 Gallager bound for $J$ independent parallel channels in order to get the tightest bound within this form (hence, the optimization is carried w.r.t. $J$ probability tilting measures). This optimization differs from the discussion in [21] where the authors choose some simple and sub-optimal tilting measures. By doing so, the authors in [21] derive bounds which are easier for numerical calculation, but the tightness of these bounds is loosened as compared to the improved bound which is combined with the $J$ optimized tilting measures (this will be exemplified in Section 7 for turbo-like ensembles).

### 4.1 Presentation of the Bound [21]

Consider a binary linear block code $\mathcal{C}$. Let $\underline{x}^m$ be the transmitted codeword and define the tilted ML metric

$$D_m(\underline{x}^{m'}, \underline{y}) \triangleq \ln\left(\frac{f_n^{(m)}(\underline{y})}{p_n(\underline{y}|\underline{x}^{m'})}\right) \qquad (45)$$

where $f_n^{(m)}(\underline{y})$ is an arbitrary function which is positive if there exists $m' \neq m$ such that $p_n(\underline{y}|\underline{x}^{m'})$ is positive. If the code is ML decoded, an error occurs if for some $m' \neq m$

$$D_m(\underline{x}^{m'}, \underline{y}) \leq D_m(\underline{x}^m, \underline{y}) .$$

As noted in [32], $D_m(\cdot, \cdot)$ is in general not computable at the receiver. It is used here as a conceptual tool to evaluate the upper bound on the ML decoding error probability. The received set $\mathcal{Y}^n$ is expressed as a union of two disjoint subsets

$$
\begin{aligned}
\mathcal{Y}^n &= \mathcal{Y}_g^n \cup \mathcal{Y}_b^n \\
\mathcal{Y}_g^n &\triangleq \{\underline{y} \in \mathcal{Y}^n : D_m(\underline{x}^m, \underline{y}) \leq nd\} \\
\mathcal{Y}_b^n &\triangleq \{\underline{y} \in \mathcal{Y}^n : D_m(\underline{x}^m, \underline{y}) > nd\}
\end{aligned}
$$

where $d$ is an arbitrary real number. The conditional ML decoding error probability can be expressed as the sum of two terms

$$P_{e|m} = \text{Prob}(\text{error}, \underline{y} \in \mathcal{Y}_b^n) + \text{Prob}(\text{error}, \underline{y} \in \mathcal{Y}_g^n)$$

which is upper bounded by

$$P_{e|m} \leq \text{Prob}(\underline{y} \in \mathcal{Y}_b^n) + \text{Prob}(\text{error}, \underline{y} \in \mathcal{Y}_g^n) \ . \tag{46}$$

We use separate bounding techniques for the two terms in (46). Applying the Chernoff bound on the first term gives

$$P_1 \triangleq \text{Prob}(\underline{y} \in \mathcal{Y}_b^n) \leq \mathbf{E}\left(e^{sW}\right), \quad s \geq 0 \tag{47}$$

where

$$W \triangleq \ln\left(\frac{f_n^{(m)}(\underline{y})}{p_n(\underline{y}|\underline{x}^m)}\right) - nd \ . \tag{48}$$

Using a combination of the union and Chernoff bounds for the second term in the RHS of (46) gives

$$
\begin{aligned}
P_2 &\triangleq \text{Prob}(\text{error}, \underline{y} \in \mathcal{Y}_g^n) \\
&= \text{Prob}\left(D_m(\underline{x}^{m'}, \underline{y}) \leq D_m(\underline{x}^m, \underline{y}) \text{ for some } m' \neq m, \ \underline{y} \in \mathcal{Y}_g^n\right) \\
&\leq \sum_{m' \neq m} \text{Prob}\left(D_m(\underline{x}^{m'}, \underline{y}) \leq D_m(\underline{x}^m, \underline{y}), \ D_m(\underline{x}^m, \underline{y}) \leq nd\right) \\
&\leq \sum_{m' \neq m} \mathbf{E}\left(\exp(tU_{m'} + rW)\right), \quad t \geq 0, r \leq 0
\end{aligned}
\tag{49}
$$

where, based on (45),

$$U_{m'} = D_m(\underline{x}^m, \underline{y}) - D_m(\underline{x}^{m'}, \underline{y}) = \ln\left(\frac{p_n(\underline{y}|\underline{x}^{m'})}{p_n(\underline{y}|\underline{x}^m)}\right) \ . \tag{50}$$

Consider a codeword of a binary linear block code $\mathcal{C}$ which is transmitted over $J$ parallel MBIOS channels. Since the conditional error probability under ML decoding does not depend on the transmitted codeword, one can assume without loss of generality that the all-zero codeword is transmitted. As before, we impose on the function $f_n^{(m)}(\underline{y})$ the restriction that it can be expressed in the product form

$$f_n^{(m)}(\underline{y}) = \prod_{j=1}^{J} \prod_{i \in \mathcal{I}(J)} f(y_i; j) \ . \tag{51}$$

For the continuation of the derivation, it is assumed that the functions $f(\cdot;j)$ are even, i.e., $f(y;j) = f(-y;j)$ for all $y \in \mathcal{Y}$. Plugging (25), (48), (50) and (51) into (47) and (49) we get

$$P_1 \leq \sum_{\underline{y}} \left\{ \prod_{j=1}^{J} \prod_{i \in \mathcal{I}(j)} \left( \frac{f(y_i;j)}{p(y_i|0;j)} \right)^s p(y_i|0;j) \right\} e^{-nsd}$$

$$= \prod_{j=1}^{J} \left\{ \left( \sum_{y \in \mathcal{Y}} p(y|0;j)^{1-s} f(y;j)^s \right)^{n_j} \right\} e^{-nsd} \quad s \geq 0 \quad (52)$$

$$P_2 \leq \sum_{\underline{y}} \sum_{m' \neq m} \prod_{j=1}^{J} \prod_{i \in \mathcal{I}(j)} \left( \frac{f(y_i;j)}{p(y_i|0;j)} \right)^r p(y_i|0;j) \left( \frac{p(y_i|0;j)}{p(y_i|x_i^{(m')};j)} \right)^t e^{-nrd}$$

$$= \sum_{h_1=0}^{n_1} \cdots \sum_{h_J=0}^{n_J} \left\{ A_{h_1,\ldots,h_J} \prod_{j=1}^{J} \left[ \sum_{y \in \mathcal{Y}} p(y|0;j)^{1-r} f(y;j)^r \left( \frac{p(y|0;j)}{p(y|1;j)} \right)^t \right]^{h_j} \right.$$

$$\left. \prod_{j=1}^{J} \left[ \sum_{y \in \mathcal{Y}} p(y|0;j)^{1-r} f(y;j)^r \right]^{n_j-h_j} \right\} e^{-nrd}, \quad t, r \leq 0 \quad (53)$$

where as before, we use the notation $n_j \triangleq |\mathcal{I}(j)|$. Optimizing the parameter $t$ gives the value in [12, Eq. (3.27)]

$$t = \frac{r-1}{2} \ . \quad (54)$$

Let us define

$$G(r;j) \triangleq \sum_{y} p(y|0;j)^{1-r} f(y;j)^r \quad (55)$$

$$Z(r;j) \triangleq \sum_{y} [p(y|0;j) p(y|1;j)]^{\frac{1-r}{2}} f(y;j)^r. \quad (56)$$

Substituting (54) into (53), combining the bounds on $P_1$ and $P_2$ in (52) and (53), and finally averaging over all possible channel assignments, we obtain

$$P_{\mathrm{e}} \leq \mathbf{E} \left[ \sum_{h=1}^{n} \sum_{\substack{0 \leq h_j \leq n_j \\ \sum h_j = h}} A_{h_1,\ldots,h_J} \prod_{j=1}^{J} [Z(r;j)]^{h_j} [G(r;j)]^{n_j-h_j} e^{-nrd} \right.$$

$$\left. + \prod_{j=1}^{J} [G(s;j)]^{n_j} e^{-nsd} \right]$$

$$= \sum_{\substack{n_j \geq 0 \\ \sum n_j = n}} \left\{ \sum_{h=1}^{n} \sum_{\substack{0 \leq h_j \leq n_j \\ \sum h_j = h}} A_{h_1,\ldots,h_J} \prod_{j=1}^{J} [Z(r;j)]^{h_j} [G(r;j)]^{n_j-h_j} e^{-nrd} \right.$$

$$\left. + \prod_{j=1}^{J} [G(s;j)]^{n_j} e^{-nsd} \right\} P_{\underline{N}}(\underline{n}) \quad , \quad \begin{array}{c} r \leq 0 \\ s \geq 0 \\ -\infty < d < \infty \end{array} \quad . \quad (57)$$

Following the same procedure for random assignments as in (30) and (31), we obtain

$$P_e \leq \sum_{h=1}^{n} \left\{ \overline{A_h} \left( \sum_{j=1}^{J} \alpha_j Z(r;j) \right)^h \left( \sum_{j=1}^{J} \alpha_j G(r;j) \right)^{n-h} \right\} e^{-nrd}$$
$$+ \left( \sum_{j=1}^{J} \alpha_j G(s;j) \right)^n e^{-nsd} . \tag{58}$$

Finally, we optimize the bound in (58) over the parameter $d$ which gives

$$P_e \leq 2^{h(\rho)} \left\{ \sum_{h=1}^{n} \overline{A_h} \left[ \sum_{j=1}^{J} \alpha_j Z(r;j) \right]^h \left[ \sum_{j=1}^{J} \alpha_j G(r;j) \right]^{n-h} \right\}^\rho \left\{ \sum_{j=1}^{J} \alpha_j G(s;j) \right\}^{n(1-\rho)} \tag{59}$$

where $r \leq 0$, $s \geq 0$, and

$$\rho \triangleq \frac{s}{s-r} \quad , \quad 0 \leq \rho \leq 1 . \tag{60}$$

The bound in (59), originally derived in [21], forms a natural generalization of the 1961 Gallager bound for parallel channels.

## 4.2 Connection to the Generalization of the DS2 Bound for Independent and Memoryless Parallel Channels

Divsalar [8], and Sason and Shamai [29, 32] discussed the connection between the DS2 bound and the 1961 Gallager bound for a single MBIOS channel. In this case, it was demonstrated that the former bound is necessarily tighter than the latter bound. At a first glance, one would expect this conclusion to be valid also for the case where the communication takes place over $J$ independent parallel MBIOS channels (where $J > 1$). We show in this section the difficulty in generalizing this conclusion to an arbitrary number of independent parallel channels, and the numerical results presented in Section 7 support the conclusion that the DS2 bound is not necessarily tighter than the 1961 Gallager bound when considering communications over an arbitrary number of independent parallel channels.

In what follows, we will see how a variation in the derivation of the Gallager bound leads to a form of the DS2 bound, up to a factor which varies between 1 and 2. To this end, we start from the point in the last section where the combination of the bounds in (52) and (53) is obtained. Rather than continuing as in the last section, we first optimize over the parameter $d$ in the sum of the bounds on $P_1$ and $P_2$ in (52) and (53), yielding that

$$P_e \leq 2^{h(\rho)} \left\{ \sum_{h=1}^{n} \sum_{\substack{h_1,\ldots,h_J \\ \sum_j h_j = h}} A_{h_1,\ldots,h_J} \prod_{j=1}^{J} V(r,t;j)^{h_j} G(r;j)^{n_j-h_j} \right\}^\rho \prod_{j=1}^{J} G(s;j)^{n_j(1-\rho)}$$

$$= 2^{h(\rho)} \left\{ \sum_{h=1}^{n} \sum_{\substack{h_1,\ldots,h_J \\ \sum_j h_j = h}} A_{h_1,\ldots,h_J} \prod_{j=1}^{J} \left[ V(r,t;j) G(s;j)^{\frac{1-\rho}{\rho}} \right]^{h_j} \right.$$
$$\left. \prod_{j=1}^{J} \left[ G(r;j) G(s;j)^{\frac{1-\rho}{\rho}} \right]^{n_j-h_j} \right\}^\rho , \quad t, r \leq 0, \ s \geq 0$$

where

$$V(r,t;j) \triangleq \sum_y p(y|0;j)^{1-r} f(y;j)^r \left(\frac{p(y|0;j)}{p(y|1;j)}\right)^t \tag{61}$$

$G(\cdot;j)$ is introduced in (55) for $j = 1, \ldots, J$, and $\rho$ is given in (60). Averaging the bound with respect to all possible channel assignments, we get for $0 \leq \rho \leq 1$

$$P_{\text{e}} \leq 2^{h(\rho)} \sum_{\substack{n_j \geq 0 \\ \sum_j n_j = n}} \left\{ \left[ \sum_{h=1}^n \sum_{\substack{h_1,\ldots,h_j \\ \sum_j h_j = h}} A_{h_1,\ldots,h_j} \prod_{j=1}^J \left[V(r,t;j)G(s;j)^{\frac{1-\rho}{\rho}}\right]^{h_j} \right. \right.$$
$$\left. \left. \prod_{j=1}^J \left[G(r;j)G(s;j)^{\frac{1-\rho}{\rho}}\right]^{n_j - h_j} \right]^\rho P_{\underline{N}}(\underline{n}) \right\}$$

$$\leq 2^{h(\rho)} \left[ \sum_{\substack{n_j \geq 0 \\ \sum_j n_j = n}} \sum_{h=1}^n \sum_{\substack{h_1,\ldots,h_j \\ \sum_j h_j = h}} A_{h_1,\ldots,h_j} P_{\underline{N}}(\underline{n}) \prod_{j=1}^J \left[V(r,t;j)G(s;j)^{\frac{1-\rho}{\rho}}\right]^{h_j} \right.$$
$$\left. \prod_{j=1}^J \left[G(r;j)G(s;j)^{\frac{1-\rho}{\rho}}\right]^{n_j - h_j} \right]^\rho \tag{62}$$

where we invoked Jensen's inequality in the last step. Following the same steps as in (28)–(31), we get

$$P_{\text{e}} \leq 2^{h(\rho)} \left[ \sum_{h=1}^n \overline{A_h} \left( \sum_{j=1}^J \alpha_j V(r,t;j) G(s;j)^{\frac{1-\rho}{\rho}} \right)^h \right.$$
$$\left. \left( \sum_{j=1}^J \alpha_j G(r;j) G(s;j)^{\frac{1-\rho}{\rho}} \right)^{n-h} \right]^\rho, \tag{63}$$

where from (54), (55), (60) and (61)

$$G(s;j) = \sum_y p(y|0;j) \left(\frac{f(y;j)}{p(y|0;j)}\right)^s$$

$$G(r;j) = \sum_y p(y|0;j) \left(\frac{f(y;j)}{p(y|0;j)}\right)^{s(1-\frac{1}{\rho})}$$

$$V(r,t;j) = \sum_y p(y|0;j) \left(\frac{f(y;j)}{p(y|0;j)}\right)^{s(1-\frac{1}{\rho})} \left(\frac{p(y|0;j)}{p(y|1;j)}\right)^t. \tag{64}$$

Setting $\lambda = -t$, and substituting in (64) the following relation between the Gallager's tilting measures and the un-normalized tilting measures in the DS2 bound

$$g(y;j) \triangleq \left(\frac{f(y;j)}{p(y|0;j)}\right)^s, \quad j = 1, 2, \ldots, J \tag{65}$$

we obtain

$$P_{\mathrm{e}} \leq 2^{h(\rho)} \left\{ \sum_{h=0}^{n} \overline{A}_h \left[ \sum_{j=1}^{J} \alpha_j \left( \sum_{y} g(y;j)^{1-\frac{1}{\rho}} p(y|0;j)^{1-\lambda} p(y|1;j)^{\lambda} \right) \right. \right.$$
$$\left. \left( \sum_{y} g(y;j) p(y|0;j) \right)^{\frac{1-\rho}{\rho}} \right]^{h} \left[ \sum_{j=1}^{J} \alpha_j \left( \sum_{y} g(y;j)^{1-\frac{1}{\rho}} p(y|0;j) \right) \right.$$
$$\left. \left. \left( \sum_{y} g(y;j) p(y|0;j) \right)^{\frac{1-\rho}{\rho}} \right]^{n-h} \right\}^{\rho}, \quad 0 \leq \rho \leq 1 \qquad (66)$$

which coincides with the form of the DS2 bound given in (33) (up to the factor $2^{h(\rho)}$ which lies between 1 and 2), for those un-normalized tilting measures $g(\cdot;j)$ such that the resulting functions $f(\cdot;j)$ in (65) are *even*.

*Discussion.* The derivation of the 1961 Gallager bound first involves the averaging of the bound in (57) over all possible channel assignments and then the optimization over the parameter $d$ in (58). To show a connection to the DS2 bound, we had first optimized over $d$ and then obtained the bound averaged over all possible channel assignments. The difference between the two approaches is that in the latter, Jensen's inequality had to be used in (62) to continue the derivation (because the expectation over all possible channel assignments was performed on an expression raised to the $\rho$-th power) which resulted in the DS2 bound, whereas in the derivation of [21], the need for Jensen's inequality was circumvented due to the linearity of the expression in (57). We note that Jensen's inequality was also used for the direct derivation of the DS2 bound in (31). In the particular case where $J = 1$, there is no need to apply Jensen's inequality in (28) and (62). By particularizing the model of parallel channels to a single MBIOS channel, the DS2 bound is tighter than the 1961 Gallager bound (as noticed in [32]) due to the following reasons:

- For the 1961 Gallager bound, it is required that $f(\cdot;j)$ be even. This requirement inhibits the optimization of $\psi(\cdot;j)$ in Section 3.2 because the optimal choice of $\psi(\cdot;j)$ given in (42) leads to functions $f(\cdot;j)$ which are not even. The exact form of $f(\cdot;j)$ which stems from the optimal choice of $\psi(\cdot;j)$ is detailed in Appendix A.1.

- The absence of the factor $2^{h(\rho)}$ (which is greater than 1) in the DS2 bound implies its superiority. Naturally, this factor is of minor importance since we are primarily interested in the exponential tightness of these bounds.

For the case where $J > 1$, it is not clear from the discussion so far which of the bounds is tighter. We note that, as in the case of $J = 1$, the optimization over the DS2 tilting measure is over a larger set of functions as compared to the 1961 Gallager tilting measure; hence, the derivation of the DS2 bound from the Gallager bound *only gives an expression of the same form and not the same upper bound* (disregarding the $2^{h(\rho)}$ constant). We will later compare the optimized DS2 bound and the optimized 1961 Gallager bound.

Although we have shown here how to obtain a form of the DS2 bound from the Gallager bound by invoking Jensen's inequality, it is insightful to use the relations connecting the DS2 bound with the Gallager bound in the derivation above without invoking Jensen's inequality, in order to obtain from the 1961 Gallager bound a tighter bound than the one in (66). This is possible if we start from the bounds in (52) and (53), optimize over all possible channel assignments and then optimize over $d$. In this way the use of Jensen's inequality is avoided, and one can obtain the following upper bound

$$P_{\mathrm{e}} \leq 2^{h(\rho)} \left\{ \sum_{h=0}^{n} \overline{A}_h \left[ \sum_{j=1}^{J} \alpha_j \left( \sum_y g(y;j)^{1-\frac{1}{\rho}} p(y|0;j)^{1-\lambda} p(y|1;j)^{\lambda} \right) \right. \right.$$

$$\left(\sum_{j=1}^{J} \alpha_j \sum_y g(y;j) p(y|0;j) \right)^{\frac{1-\rho}{\rho}} \right]^h \left[ \sum_{j=1}^{J} \alpha_j \left( \sum_y g(y;j)^{1-\frac{1}{\rho}} p(y|0;j) \right) \right.$$

$$\left. \left. \left( \sum_{j=1}^{J} \alpha_j \sum_y g(y;j) p(y|0;j) \right)^{\frac{1-\rho}{\rho}} \right]^{n-h} \right\}^{\rho}. \tag{67}$$

Some technical details related to the derivation of this bound are provided in Appendix A.2. This upper bound is clearly tighter than the one in (66). The reader should note, however, that these two bounds coincide whenever the expression $\sum_y g(y;j) p(y|0;j)$ is independent of $j$.

## 4.3  Optimized Tilting Measures for the Generalized 1961 Gallager Bound

We derive in this section optimized tilting measures for the 1961 Gallager bound. These optimized tilting measures are derived for random coding, and for the case of constant Hamming weight codes. The 1961 Gallager bound will be used later in conjunction with these optimized tilting measures in order to get an upper bound on the decoding error probability of an arbitrary binary linear block code. To this end, such a code is partitioned to constant Hamming weight subcodes (where each one of them also includes the all-zero codeword), and a union bound is used in conjunction with the calculation of the conditional error probability of each subcode, given that the all-zero codeword is transmitted. Using these optimized tilting measures improves the tightness of the resulting bound, as exemplified in the continuation of this paper.

### 4.3.1  Tilting Measures for Random Codes

Consider the ensemble of fully random binary block codes of length $n$. Substituting the appropriate weight enumerator (given in (15)) into (58), we get

$$\overline{P_{\mathrm{e}}} \leq 2^{-n(1-R)} \left\{ \frac{1}{2} \sum_{j=1}^{J} \alpha_j \sum_y \left[ p(y|0;j)^{\frac{1-r}{2}} + p(y|1;j)^{\frac{1-r}{2}} \right]^2 f(y;j)^r \right\}^n e^{-nrd}$$

$$+ \left\{ \frac{1}{2} \sum_{j=1}^{J} \alpha_j \sum_y \left( p(y|0;j)^{1-s} + p(y|1;j)^{1-s} \right) f(y;j)^s \right\}^n e^{-nsd} \quad , \quad \begin{array}{c} r \leq 0 \\ s \geq 0 \\ -\infty < d < +\infty \end{array} \tag{68}$$

where we rely on (55) and (56), use the symmetry of the channels and the fact that we require the functions $f(\cdot;j)$ $(j = 1, \ldots, J)$ to be even. To optimize (68) over all possible tilting measures, we apply calculus of variations. This procedure gives the following equation:

$$\sum_{j=1}^{J} \alpha_j \left( p(y|0;j)^{\frac{1-r}{2}} + p(y|1;j)^{\frac{1-r}{2}} \right)^2 f(y;j)^{r-1}$$

$$-L \sum_{j=1}^{J} \alpha_j \left( p(y|0;j)^{1-s} + p(y|1;j)^{1-s} \right)^2 f(y;j)^{s-1} = 0 \quad \forall y.$$

where $L \in \mathbb{R}$. This equation is satisfied for functions which are given in the form

$$f(y;j) = K \left\{ \frac{\left( p(y|0;j)^{\frac{1-r}{2}} + p(y|1;j)^{\frac{1-r}{2}} \right)^2}{p(y|0;j)^{1-s} + p(y|1;j)^{1-s}} \right\}^{\frac{1}{s-r}} \quad K \in \mathbb{R}. \tag{69}$$

This forms a natural generalization of the tilting measure given in [12, Eq. (3.41)] for a single MBIOS channel. We note that the scaling factor $K$ may be omitted as it cancels out when we substitute (69) in (59).

### 4.3.2  Tilting Measures for Constant Hamming Weight Codes

The distance spectrum of a constant Hamming weight code is given by

$$A_{h'} = \begin{cases} 1, & \text{if } h' = 0 \\ A_h, & \text{if } h' = h \\ 0, & \text{otherwise} \end{cases} \tag{70}$$

Substituting this into (59) and using the symmetry of the component channels and the fact that the tilting measures $f(\cdot;j)$ are required to be even, we get

$$P_e \leq 2^{h(\rho)} A_h^\rho \left\{ \sum_{j=1}^J \alpha_j \sum_y [p(y|0;j)p(y|1;j)]^{\frac{1-r}{2}} f(y;j)^r \right\}^{h\rho}$$

$$\cdot \left\{ \sum_{j=1}^J \frac{\alpha_j}{2} \sum_y \left[ p(y|0;j)^{1-r} + p(y|1;j)^{1-r} \right] f(y;j)^r \right\}^{(n-h)\rho}$$

$$\cdot \left\{ \sum_{j=1}^J \frac{\alpha_j}{2} \sum_y \left[ p(y|0;j)^{1-s} + p(y|1;j)^{1-s} \right] f(y;j)^s \right\}^{n(1-\rho)},$$

$$r \leq 0, \ s \geq 0, \ \rho = \frac{s}{s-r}. \tag{71}$$

Applying calculus of variations to (71) yields (see Appendix A.3 for some additional details) that the following condition should be satisfied for all values of $y \in \mathcal{Y}$:

$$\sum_{j=1}^J \alpha_j \left\{ \left[ p(y|0;j)^{1-s} + p(y|1;j)^{1-s} \right] f(y;j)^{s-r} + K_1 \left[ p(y|0;j)p(y|1;j) \right]^{\frac{1-r}{2}} \right. \tag{72}$$

$$\left. + K_2 \left[ p(y|0;j)^{1-r} + p(y|1;j)^{1-r} \right] \right\} = 0$$

where $K_1, K_2 \in \mathbb{R}$. This condition is satisfied if we require

$$\left[ p(y|0;j)^{1-s} + p(y|1;j)^{1-s} \right] f(y;j)^{s-r} + K_1 \left[ p(y|0;j)p(y|1;j) \right]^{\frac{1-r}{2}}$$
$$+ K_2 \left[ p(y|0;j)^{1-r} + p(y|1;j)^{1-r} \right] \equiv 0, \quad \forall y \in \mathcal{Y}, \ j = 1, \ldots, J.$$

The optimized tilting measures can therefore be expressed in the form

$$f(y;j) = \left\{ \frac{c_1 \left( p(y|0;j)^{\frac{1-s(1-\rho^{-1})}{2}} + p(y|1;j)^{\frac{1-s(1-\rho^{-1})}{2}} \right)^2}{p(y|0;j)^{1-s} + p(y|1;j)^{1-s}} + \frac{d_1 \left( p(y|0;j)^{1-s(1-\rho^{-1})} + p(y|1;j)^{1-s(1-\rho^{-1})} \right)}{p(y|0;j)^{1-s} + p(y|1;j)^{1-s}} \right\}^{\frac{\rho}{s}}, \quad c_1, d_1 \in \mathbb{R}, s \geq 0, 0 \leq \rho \leq 1 \quad (73)$$

where we have used (60). This form is identical to the optimal tilting measure for random codes if we set $d_1 = 0$. It is possible to scale the parameters $c_1$ and $d_1$ without affecting the 1961 Gallager bound (i.e., the ratio $\frac{c_1}{d_1}$ cancels out when we substitute (73) in (59)). Furthermore, we note that regardless of the values of $c_1$ and $d_1$, the resulting tilting measures are even functions, as required in the derivation of the 1961 Gallager bound.

For the simplicity of the optimization, we wish to reduce the infinite intervals in (73) to finite ones. It is shown in [27, Appendix A] that the optimization of the parameter $s$ can be reduced to the interval $[0, 1]$ without loosening the tightness of the bound. Furthermore, the substitution $c \triangleq \frac{c_1 + 2d_1}{2c_1 + 3d_1}$, as suggested in [27, Appendix B], enables one to express the optimized tilting measure in (73) using an equivalent form where the new parameter $c$ ranges in the interval $[0, 1]$. The numerical optimization of the bound in (73) is therefore taken over the range of parameters $0 \leq \rho \leq 1$, $0 \leq s \leq 1$, $0 \leq c \leq 1$. Based on the calculations in [27, Appendices A, B], the functions $f(\cdot; j)$ get the equivalent form

$$f(y;j) = \left\{ \frac{(1-c) \left( p(y|0;j)^{\frac{1-s(1-\rho^{-1})}{2}} - p(y|1;j)^{\frac{1-s(1-\rho^{-1})}{2}} \right)^2}{p(y|0;j)^{1-s} + p(y|1;j)^{1-s}} + \frac{2c \left( p(y|0;j) p(y|1;j) \right)^{\frac{1-s(1-\rho^{-1})}{2}}}{p(y|0;j)^{1-s} + p(y|1;j)^{1-s}} \right\}^{\frac{\rho}{s}}, \quad (\rho, s, c) \in [0,1]^3. \quad (74)$$

By reducing the optimization of the three parameters over the unit cube, the complexity of the numerical process is reduced to an acceptable level.

### 4.4 Statement of the Main Result Derived in Section 4

The analysis in this section leads to the following theorem:

**Theorem 2 (Generalized 1961 Gallager bound for independent parallel MBIOS channels with optimized tilting measures).** Consider the transmission of binary linear block codes (or ensembles) over a set of $J$ independent parallel MBIOS channels. Following the notation in Theorem 1, the generalization of the 1961 Gallager bound in (59) provides an upper bound on the ML decoding error probability when the bound is taken over the whole code (as originally derived in [21]). By partitioning the code into constant Hamming-weight subcodes, the generalized 1961 Gallager bound on the conditional ML decoding error probability of an arbitrary subcode (given the all-zero codeword is transmitted) is provided by (71), and (34) forms an upper bound on the block error probability of the whole code (or ensemble). For an arbitrary constant Hamming weight subcode, the optimized set of non-negative and even functions $\{f(\cdot; j)\}_{j=1}^{J}$ which attains the minimal value of the conditional bound in (71), is given by (74); this set of functions is subject to a three-parameter optimization over a cube of unit length.

# 5 Special Cases of the Generalized DS2 Bound for Independent Parallel Channels

In this section, we rely on the generalized DS2 bound for independent parallel MBIOS channels, as presented in Section 3.1, and apply it in order to re-derive some of the bounds which were originally derived by Liu et al. [21]. The derivation in [21] is based on the 1961 Gallager bound from Section 4.1, and the authors choose particular and sub-optimal tilting measures in order to get closed form bounds (in contrast to the optimized tilting in Section 4.3 which lead to more complicated bounds in terms of their numerical computation). In this section, we follow the same approach in order to re-derive some of their bounds as particular cases of the generalized DS2 bound (i.e., we choose some particular tilting measures rather than the optimized ones in Section 3.2). In some cases, we re-derive the bounds from [21] as special cases of the generalized DS2 bound, or alternatively, obtain some modified bounds as compared to [21].

## 5.1 Union-Bhattacharyya Bound in Exponential Form

As in the case of a single channel, it is a special case of both the DS2 and Gallager bounds. By substituting $r = 0$ in the Gallager bound or $\rho = 1, \lambda = 0.5$ in the DS2 bound, we get

$$\overline{P_{\text{e}}} \leq \sum_{h=1}^{n} \overline{A}_h \gamma^h$$

where $\gamma$ is given by (3) and denotes the average Bhattacharyya parameter of $J$ independent parallel channels. Note that this bound is given in exponential form, i.e., as in the single channel case, it doesn't use the exact expression for the pairwise error probability between two codewords of Hamming distance $h$. For the case of the AWGN, a tighter version which uses the $Q$-function to express the exact pairwise error probability is presented in Appendix C.

## 5.2 The Sphere Bound for Parallel AWGN Channels

The simplified sphere bound is an upper bound on the decoding error probability for the binary-input AWGN channel. In [21], the authors have obtained a parallel-channel version of the sphere bound by making the substitution $f(y;j) = \frac{1}{\sqrt{2\pi}}$ in the 1961 Gallager bound. We will show that this version is also a special case of the parallel-channel DS2 bound. By using the relation (65), between Gallager's tilting measure and the un-normalized DS2 tilting measure, we get

$$g(y;j) = \left(\frac{f(y;j)}{p(y|0;j)}\right)^s = \exp\left(\frac{s(y + \sqrt{2\nu_j})^2}{2}\right)$$

so that

$$\int_{-\infty}^{+\infty} g(y;j) p(y|0;j) \, dy = \frac{1}{\sqrt{1-s}}$$

$$\int_{-\infty}^{+\infty} g(y;j)^{1-\frac{1}{\rho}} p(y|0;j) \, dy = \frac{1}{\sqrt{1 - s\left(1 - \frac{1}{\rho}\right)}}$$

$$\int_{-\infty}^{+\infty} g(y;j)^{1-\frac{1}{\rho}} p(y|0;j)^{1-\lambda} p(y|1;j)^\lambda \, dy = \frac{e^{\nu_j\left(1 - s\left(1 - \frac{1}{\rho}\right)\right)}}{\sqrt{1 - s\left(1 - \frac{1}{\rho}\right)}}.$$

By introducing the two new parameters $\beta = 1 - s\left(1 - \frac{1}{\rho}\right)$ and $\lambda = \beta/2$ we get

$$\int_{-\infty}^{+\infty} g(y;j)p(y|0;j)dy = \sqrt{\frac{1-\rho}{1-\beta\rho}}$$
$$\int_{-\infty}^{+\infty} g(y;j)^{1-\frac{1}{\rho}}p(y|0;j)dy = \beta^{-\frac{1}{2}} \quad (75)$$
$$\int_{-\infty}^{+\infty} g(y;j)^{1-\frac{1}{\rho}}p(y|0;j)^{1-\lambda}p(y|1;j)^{\lambda}dy = \frac{\gamma_j^{\beta}}{\sqrt{\beta}}, \quad \gamma_j \triangleq e^{\nu_j}.$$

Next, by plugging (75) into (33), we get

$$\overline{P_e} \leq \left\{\sum_{h=0}^{n} \overline{A_h}\left(\sum_{j=1}^{J} \alpha_j \gamma_j^{\beta}\right)^h \beta^{-\frac{n}{2}}\right\}^{\rho} \left(\frac{1-\rho}{1-\beta\rho}\right)^{\frac{n(1-\rho)}{2}}, \quad \begin{array}{c} 0 \leq \rho \leq 1 \\ 1 \leq \beta \leq \frac{1}{\rho} \end{array}. \quad (76)$$

This bound is identical to the parallel-channel simplified sphere bound in [21, Eq. (24)], except that it provides a slight improvement due to the absence of the factor $2^{h(\rho)}$ which appears in [21, Eq. (24)] (a factor bounded between 1 and 2).

We observe that in (75), $\int_y g(y;j)p(y|0;j)dy$ is independent of $j$, so the fact that the expressions in (76) and [21, Eq. (24)] coincide is not surprising in light of the discussion at the end of Section 4.2.

### 5.3 Generalizations of the Shulman-Feder Bound for Parallel Channels

In this sub-section, we present two generalizations of the Shulman and Feder (SF) bound, where both bounds apply to independent parallel channels. The first bound was previously obtained by Liu et al. [21] as a special case of the generalization of the 1961 Gallager bound, and the second bound follows as a particular case of the generalized DS2 bound for independent parallel channels.

By substituting in (59) the tilting measure and the parameters (see [21, Eq. (28)])

$$f(y;j) = \left(\frac{1}{2}p(y|0;j)^{\frac{1}{1+\rho}} + \frac{1}{2}p(y|1;j)^{\frac{1}{1+\rho}}\right)^{1+\rho}$$
$$r = -\frac{1-\rho}{1+\rho}, \quad s = \frac{\rho}{1+\rho}, \quad 0 \leq \rho \leq 1 \quad (77)$$

straightforward calculations for MBIOS channels give the following bound which was originally introduced in [21, Lemma 2]:

$$\overline{P_e} \leq 2^{h(\rho)} 2^{nR\rho} \left(\max_{1 \leq h \leq n} \frac{\overline{A_h}}{2^{-n(1-R)}\binom{n}{h}}\right)^{\rho} \left\{\sum_{j=1}^{J} \alpha_j \left(\sum_y \frac{1}{2}p(y|0;j)^{\frac{1}{1+\rho}} + \frac{1}{2}p(y|1;j)^{\frac{1}{1+\rho}}\right)^{1+\rho}\right\}^n. \quad (78)$$

Considering the DS2 bound, it is possible to start from Eq. (33) and take the maximum distance spectrum term out of the sum. This gives the bound

$$\overline{P_e} \leq 2^{-n(1-R)\rho} \left(\max_{1 \leq h \leq n} \frac{\overline{A_h}}{2^{-n(1-R)}\binom{n}{h}}\right)^{\rho} \left\{\sum_{j=1}^{J} \alpha_j \left[\sum_y g(y;j)p(y|0;j)\right]^{\frac{1-\rho}{\rho}}\right.$$
$$\left. \cdot \left[\sum_y p(y|0;j)g(y;j)^{1-\frac{1}{\rho}}\left(1 + \left(\frac{p(y|1;j)}{p(y|0;j)}\right)^{\lambda}\right)\right]\right\}^{n\rho}, \quad 0 \leq \rho \leq 1. \quad (79)$$

Using the $J$ un-normalized tilting measures

$$g(y;j) = \left[\frac{1}{2} p(y|0;j)^{\frac{1}{1+\rho}} + \frac{1}{2} p(y|1;j)^{\frac{1}{1+\rho}}\right]^{\rho} p(y|0;j)^{-\frac{\rho}{1+\rho}} \quad , \quad j = 1, 2, \ldots, J \quad (80)$$

and setting $\lambda = \frac{1}{1+\rho}$ in (79), gives the following bound due to the symmetry at the channel outputs:

$$\overline{P_{\text{e}}} \leq 2^{nR\rho} \left(\max_{1 \leq h \leq n} \frac{\overline{A}_h}{2^{-n(1-R)}\binom{n}{h}}\right)^{\rho}$$

$$\left\{\sum_{j=1}^{J} \alpha_j \left[\left(\sum_y \frac{1}{2}p(y|0;j)^{\frac{1}{1+\rho}} + \frac{1}{2}p(y|1;j)^{\frac{1}{1+\rho}}\right)^{1+\rho}\right]^{\frac{1}{\rho}}\right\}^{n\rho}, \quad 0 \leq \rho \leq 1 \quad (81)$$

which forms another possible generalization of the SF bound for independent parallel channels, where the latter variation follows from the generalized DS2 bound. Clearly, unless $J = 1$ (referring to the case of a single MBIOS channel), this bound is exponentially looser than the one in (78). The fact that the bound in (81) is exponentially looser than the bound in (78) follows from the use of Jensen's inequality for the derivation of the generalized DS2 bound (see the move from (28) to (29)). The tilting measures $g(\cdot;j)$ and $f(\cdot;j)$ ($j = 1, \ldots, J$) in (80) and (77), respectively, satisfy the relation in (65). For the case where $J = 1$, these two bounds coincide (up to the $2^{h(\rho)}$ term which makes the 1961 Gallager bound at most twice larger than the DS2 bound), and one gets that this particular case of DS2 bound is identical to the SF bound, as originally observed in [32].

## 5.4 Modified Shulman-Feder Bound for Independent Parallel Channels

It is apparent from the form of the SF bound that its exponential tightness depends on the quantity

$$\max_{1 \leq h \leq n} \frac{\overline{A}_h}{2^{-n(1-R)}\binom{n}{h}} \quad (82)$$

which measures the maximal ratio of the distance spectrum of the considered binary linear block code (or ensemble) and the average distance spectrum of fully random block codes with the same rate and block length. One can observe from Fig. 3 (see p. 9) that this ratio may be quite large for a non-negligible portion of the normalized Hamming weights, thus undermining the tightness of the SF bound. The idea of the modified Shulman-Feder (MSF) bound is to split the set of non-zero normalized Hamming weights $\Psi_n \triangleq \{\frac{1}{n}, \frac{2}{n}, \ldots, 1\}$ into two disjoint subsets $\Psi_n^+$ and $\Psi_n^-$ where the union bound is used for the codewords with normalized Hamming weights within the set $\Psi_n^+$, and the SF bound is used for the remaining codewords. This concept was originally applied to the ML analysis of ensembles of LDPC codes by Miller and Burshtein [24]. Typically, the set $\Psi_n^+$ consists of low and high Hamming weights, where the ratio in (82) between the distance spectra and the binomial distribution appear to be quite large for typical code ensembles of linear codes; the set $\Psi_n^-$ is the complementary set which includes medium values of the normalized Hamming weight. The MSF bound for a given partitioning $\Psi_n^-, \Psi_n^+$ is introduced in [21, Lemma 3], and gets the form

$$\overline{P_{\text{e}}} \leq \sum_{h: \frac{h}{n} \in \Psi_n^+} \overline{A}_h \gamma^h +$$

$$2^{h(\rho)} 2^{nR\rho} \left(\max_{h: \frac{h}{n} \in \Psi_n^-} \frac{\overline{A}_h}{2^{-n(1-R)}\binom{n}{h}}\right)^{\rho} \left\{\sum_{j=1}^{J} \alpha_j \left(\sum_y \frac{1}{2}p(y|0;j)^{\frac{1}{1+\rho}} + \frac{1}{2}p(y|1;j)^{\frac{1}{1+\rho}}\right)^{1+\rho}\right\}^{n} \quad (83)$$

where $\gamma$ is introduced in (3), and $0 \leq \rho \leq 1$. Liu et al. prove that in the limit where the block length tends to infinity, the optimal partitioning of the set of non-zero normalized Hamming weights to two disjoint subsets $\Psi_n^-$ and $\Psi_n^+$ is given by (see [21, Eq. (42)])

$$\delta \in \begin{cases} \Psi_n^+ & \text{if } -\delta \ln \gamma \geq H(\delta) + (\overline{I} - 1) \ln 2 \\ \Psi_n^- & \text{otherwise} \end{cases} \quad (84)$$

where

$$\overline{I} \triangleq \sum_{j=1}^{J} \frac{\alpha_j}{2} \sum_{x \in \{-1,1\}} \sum_y p(y|x;j) \log_2 \frac{p(y|x;j)}{1/2 \sum_{x' \in \{-1,1\}} p(y|x';j)}$$

designates the average mutual information under the assumption of equiprobable binary inputs. Note that for finite block lengths, even with the same partitioning as above, the first term in the RHS of (83) can be tightened by replacing the Bhattacharyya bound with the exact expression for the average pairwise error probability between two codewords of Hamming distance $h$. Referring to parallel binary-input AWGN channels, the exact pairwise error probability is given in (C.5), thus providing the following tightened upper bound:

$$\overline{P_e} \leq \frac{1}{\pi} \int_0^{\frac{\pi}{2}} \sum_{h: \frac{h}{n} \in \Psi_n^+} \overline{A_h} \left[ \sum_{j=1}^{J} \alpha_j e^{-\frac{\nu_j}{\sin^2 \theta}} \right]^h d\theta$$

$$+ 2^{h(\rho)} 2^{nR\rho} \left( \max_{h: \frac{h}{n} \in \Psi_n^-} \frac{\overline{A_h}}{2^{-n(1-R)} \binom{n}{h}} \right)^\rho \left\{ \sum_{j=1}^{J} \alpha_j \left( \sum_y \frac{1}{2} p(y|0;j)^{\frac{1}{1+\rho}} + \frac{1}{2} p(y|1;j)^{\frac{1}{1+\rho}} \right)^{1+\rho} \right\}^n. \quad (85)$$

*On the selection of a suitable partitioning of the set $\Psi_n$ in (85):* The asymptotic partitioning suggested in (84) typically yields that the union bound is used for low and high values of normalized Hamming weights; for these values, the distance spectrum of ensembles of turbo-like codes deviates considerably from the binomial distribution (referring to the ensemble of fully random block codes of the same block length and rate). Let $\delta_l$ and $\delta_r$ be the smallest and largest normalized Hamming weights, respectively, referring to the range of values ($\delta$) in (84) for which $\Psi_n^- \triangleq \{\delta_l, \delta_l + \frac{1}{n}, \ldots, \delta_r\}$, and $\Psi_n^+ \triangleq \{\frac{1}{n}, \frac{2}{n}, \ldots, \delta_l - \frac{1}{n}\} \cup \{\delta_r + \frac{1}{n}, \delta_r + \frac{2}{n}, \ldots, 1\}$ be the complementary set of normalized Hamming weights. The subsets $\Psi_n^+$ and $\Psi_n^-$ refer to the discrete values of normalized Hamming weights for which the union bound in its exponential form is superior to the SF bound and vice versa, respectively (see (83)). Our numerical experiments show that for finite-length codes (especially, for codes of small and moderate block lengths), this choice of $\delta_l$ and $\delta_r$ often happens to be sub-optimal in the sense of minimizing the overall upper bounds in (83) and (85). This happens because for $\delta = \delta_l$ (which is the left endpoint of the interval for which the SF bound is calculated), the ratio of the average distance spectrum of the considered ensemble and the one which corresponds to fully random block codes is rather large, so the second term in the RHS of (83) and (85) corresponding to the contribution of the SF bound to the overall bound is considerably larger than the first term which refers to the union bound. Therefore, for finite-length codes, the following algorithm is proposed to optimize the partition $\Psi_n = \Psi_n^+ \cup \Psi_n^-$:

1. Select initial values $\delta_{l_0}$ and $\delta_{r_0}$ (for $\delta_l$ and $\delta_r$) via (84). If there are less than two solutions to the equation $-\delta \ln \gamma = H(\delta) + (\overline{I} - 1) \ln 2$, select $\Psi_n^+ = \Psi_n$, $\Psi_n^- = \phi$ as the empty set.

2. Optimize the value of $\delta_l$ by performing a linear search in the range $[\delta_{l_0}, \delta_{r_0}]$ and finding the value of $\delta_l$ which minimizes the overall bound in the RHS of (85).

This algorithm is applied to the calculation of the LMSF bound for finite-length codes (see, e.g., Fig. 4(b) in p. 55).

Clearly, an alternative version of the MSF bound can be obtained from the generalized DS2 bound for parallel channels. In light of the discussion in Section 5.3, this version is also expected to be looser than the one in (83). We address the MSF bound in Section 6, where for various ensembles of turbo-like codes, its tightness is compared with that of the generalized DS2 and Gallager bounds.

# 6 Inner Bounds on Attainable Channel Regions for Ensembles of Good Binary Linear Codes Transmitted over Parallel Channels

In this section, we consider inner bounds on the attainable channel regions for ensembles of good binary linear codes (e.g., turbo-like codes) whose transmission takes place over independent parallel channels. The computation of these regions follows from the upper bounds on the ML decoding error probability we have obtained in Sections 3 and 4 (see Theorems 1 and 2), referring here to the asymptotic case where we let the block length tend to infinity.

Let us consider an ensemble of binary linear codes, and assume that the codewords of each code are transmitted with equal probability. A $J$-tuple of transition probabilities characterizing a parallel channel is said to be an *attainable channel point* with respect to a code ensemble $\mathcal{C}$ if the average ML decoding error probability vanishes as we let the block length tend to infinity. The *attainable channel region* of an ensemble whose transmission takes place over parallel channels is defined as the closure of the set of attainable channel points. We will focus here on the case where each of the $J$ independent parallel channels can be described by a single real parameter, i.e., the attainable channel region is a subset of $\mathbb{R}^J$; the boundary of the attainable region is called the *noise boundary* of the channel. Since the exact decoding error probability under ML decoding is in general unknown, then similarly to [21], we evaluate inner bounds on the attainable channel regions whose calculation is based on upper bounds on the ML decoding error probability.

In [21, Section 4], Liu et al. have used special cases of the 1961 Gallager bound to derive a simplified algorithm for calculating inner bounds on attainable channel regions. As compared to the bounds introduced in [21], the improvement in the tightness of the bounds presented in Theorems 1 and 2 is expected to enlarge the corresponding inner bounds on the attainable channel regions. Our numerical results referring to inner bounds on attainable channel regions are based on the following theorem:

**Theorem 3 (Inner bounds on the attainable channel regions for parallel channels).** Let us assume that the transmission of a sequence of binary linear block codes (or ensembles) $\{[\mathcal{C}(n)]\}$ takes place over a set of $J$ parallel MBIOS channels. Assume that the bits are randomly assigned to these channels, so that every bit is transmitted over a single channel and the a-priori probability for transmitting a bit over the $j$-th channel is $\alpha_j$ (where $\sum_{j=1}^{J} \alpha_j = 1$ and $\alpha_j \geq 0$ for $j \in \{1, \ldots, J\}$). Let $\{A_h^{[\mathcal{C}(n)]}\}$ designate the (average) distance spectrum of the sequence of codes (or ensembles), $r^{[\mathcal{C}]}(\delta)$ designate the asymptotic exponent of the (average) distance spectrum, and

$$\gamma_j \triangleq \sum_{y \in \mathcal{Y}} \sqrt{p(y|0;j)p(y|1;j)}, \quad j \in \{1, \ldots, J\}$$

designate the Bhattachryya constants of the channels. Assume that the following conditions hold:

1.
$$\inf_{\delta_0 < \delta \leq 1} E^{\text{DS2}}(\delta) > 0, \quad \forall \, \delta_0 \in (0, 1) \tag{86}$$

where, for $0 < \delta \leq 1$, $E^{\text{DS2}}(\delta)$ is calculated from (37) by maximizing w.r.t. $\lambda$, $\rho$ ($\lambda \geq 0$ and $0 \leq \rho \leq 1$) and the probability tilting measures $\{\psi(\cdot; j)\}_{j=1}^{J}$.

2. The inequality

$$\limsup_{\delta \to 0} \frac{r^{[\mathcal{C}]}(\delta)}{\delta} < -\ln\left(\sum_{j=1}^{J} \alpha_j \gamma_j\right) \tag{87}$$

is satisfied, where the sum inside the logarithm designates the average Bhattacharrya constant over the $J$ parallel channels, and $r^{[\mathcal{C}]}(\delta)$ designates the asymptotic growth rate of the distance spectrum as defined in (14).

3. There exists a sequence $\{D_n\}$ of natural numbers tending to infinity with increasing $n$ so that

$$\limsup_{n \to \infty} \sum_{h=1}^{D_n} A_h^{[\mathcal{C}(n)]} = 0 \tag{88}$$

4. The normalized exponent of the distance spectrum $r^{[\mathcal{C}(n)]}$ converges uniformly in $\delta \in [0,1]$ to its asymptotic limit.

Then, the $J$-tuple vector of parameters characterizing these channels lies within the attainable channel region under ML decoding.

*Proof.* The reader is referred to Appendix B. □

*Discussion:* We note that conditions 3 and 4 in Theorem 3 are similar to the last two conditions in [20, Theorem 2.3]. Condition 2 above happens to be a natural generalization of the second condition in [20, Theorem 2.3], thus generalizing the single channel case to a set of parallel channels. The distinction between [20, Theorem 2.3] which relates to typical-pairs decoding over a single channel and the statement in Theorem 3 for ML decoding over a set of independent parallel channels lies mainly in the first condition of both theorems.

A similar result which involves the generalized 1961 Gallager bound for parallel channels can be proven in the same way by replacing the first condition with an equivalent relation involving the exponent of the 1961 Gallager bound maximized over its parameters, instead of the error exponent of the DS2 bound.

The difference of our results from those presented in [21] stems from the fact that we rely here on the generalized DS2 bound and the 1961 Gallager bound with their related *optimized tilting measures*, and not on particular cases of the latter bound. These optimizations which are carried over the tilting measures of both bounds provide tighter bounds as compared to the bounds introduced in [21, Sections 4 and 5] which follow from the particular choices of the tilting measures for the generalized 1961 Gallager bound.

We later exemplify our inner bounds on the attainable channel regions for ensembles of accumulate-based codes whose transmission takes place over parallel AWGN channels. The simplest ensemble we consider is the ensemble of uniformly interleaved and non-systematic repeat-accumulate (NSRA) codes with $q \geq 3$ repetitions. It is shown in [7, Section 5] that the third condition in Theorem 3 is satisfied for this ensemble, and more explicitly

$$\sum_{h=1}^{D_n} A_h^{[\mathcal{C}(n)]} = O\left(\frac{1}{n}\right)$$

where $D_n = O\left(\ln(n)\right)$ (so the sequence $\{D_n\}$ tends to infinity logarithmically with $n$). Based on the calculations of the distance spectrum of this ensemble (see [7, Section 4]), the fourth condition in Theorem 3 is also satisfied. We note that for this ensemble, the asymptotic growth rate of the distance spectrum satisfies

$$r^{[\mathcal{C}]}(0) = 0, \quad \limsup_{\delta \to 0} \frac{r^{[\mathcal{C}]}(\delta)}{\delta} = \left.\frac{d}{d\delta} r^{[\mathcal{C}]}(\delta)\right|_{\delta=0} = 0.$$

Hence, inequality (87) in Theorem 3 (i.e., the second condition in this theorem) is also satisfied for this ensemble (since the RHS of (87) is always positive). Hence, the fulfillment of all the conditions in Theorem 3 for this ensemble requires to check under which conditions the error exponent is strictly positive (see the condition in (86)).

As a second example, for the Gallager ensembles of regular $(n, j, k)$ LDPC codes, the second, third and fourth conditions are also satisfied for the case where $j \geq 3$. Under this assumption, the minimum distance even grows linearly with the block length (see [12, Section 2.2]), so the LHS of (87) becomes negative.

We make use of the fulfillment of the condition in (87) for regular NSRA codes and some other variants of accumulate-based codes later in Section 7.2.

It is important to note that the low Hamming weight codewords which are addressed by the requirement in (88) may yield that the error probability under ML decoding does not necessarily vanish exponentially with the block length (see, e.g., [24, Theorems 3 and 4] and [7, Section 5], where the ML decoding error probability of the considered ensembles of turbo-like codes vanish asymptotically like the inverse of a polynomial of the block length).

Similarly to [20], the condition in (86) almost implies the fulfillment of inequality (87). However, the former condition in (86) may imply the fulfillment of the latter condition in (87) with equality. This note is clarified in the proof of Theorem 3 (see Appendix B).

# 7 Performance Bounds for Turbo-Like Ensembles over Parallel Channels: Moderate Block Lengths and the Asymptotic Case

In this section, we exemplify the performance bounds derived in this paper for various ensembles of turbo-like codes whose transmission is assumed to take place over parallel BIAWGN channels. We also compare the bounds to those introduced in [21], showing the superiority of the new bounds introduced in Sections 3 and 4. As mentioned before, the superiority of the generalized 1961 Gallager bound in Section 4 over the LMSF bound from [21] is attributed to the optimizations of the related tilting measures over each of the individual channels.

We especially focus on ensembles of uniformly interleaved repeat-accumulate (RA) codes and accumulate-repeat-accumulate (ARA) codes. These codes, originally introduced by Divsalar et al. [1, 7], are attractive since they possess low encoding and decoding complexity under iterative decoding and remarkable improvement in performance over classical algebraic codes. For independent parallel channels, we study their theoretical performance under ML decoding. The section considers both finite-length analysis and asymptotic analysis. In the former case, we present upper bounds on the ML decoding error probability, and in the latter case, we consider inner bounds on the attainable channel regions of these ensembles and study the gap to the capacity region. In order to assess the tightness of the bounds for ensembles of relatively short block lengths, we compare the upper bounds under optimal ML decoding with computer simulations under (sub-optimal) itera-tive decoding. In some cases, the upper bounds under ML decoding are more pessimistic than the experimental results of the iterative decoder, thus indicating that there is still room for improving the tightness of the new bounds.

The structure of this section is as follows. Section 7.1 exemplifies performance bounds for ensembles of short to moderate block length by focusing on a uniformly interleaved ensemble of turbo codes, comparing various bounds on the bit error probability under ML decoding and compare the results with computer simulation of the Log-MAP iterative decoding. Section 7.2 focuses on performance bounds for repeat-accumulate codes and their recent variations which are attractive due to their remarkable performance and low encoding and decoding complexity under iterative decoding algorithms. This sub-section analyzes the performance under optimal ML decoding, assuming the communications takes place over parallel channels. The inner bounds on the attainable channel regions whose calculations are based on Theorem 3 considerably extend the channel region which corresponds to the cutoff rate, and outperform previously reported bounds. We conclude the discussion in this section with practical considerations related to efficient implementations of the generalized DS2 and 1961 Gallager bounds for parallel channels, thus aiming to reduce the computational complexity related to the evaluation of these bounds (see Section 7.3).

## 7.1 Performance Bounds for Uniformly Interleaved Turbo Codes

In this sub-section, we exemplify the tightness of the new bounds by referring to an ensemble of uniformly interleaved turbo codes, and comparing the upper bounds on the bit error probability under ML decoding with computer simulations of an iterative decoder. The bounds for turbo code ensembles refer to parallel BIAWGN channels. The reader is referred to [19] which introduces coding theorems for turbo code ensembles under ML decoding, assuming that the transmission takes place over a single MBIOS channel (i.e., $J = 1$ in our setting).

Fig. 4 compares upper bounds on the bit error probability of the ensemble of uniformly interleaved turbo codes of rate $R = \frac{1}{3}$ bits per channel use (see Fig. 4(a)). The calculation of the average distance spectrum and IOWE of this ensemble is performed by calculating the IOWE of the constituent codes which are recursive systematic convolutional codes (to this end, we rely on the general approach provided in [23] for the calculation of the IOWE of convolutional codes), and finally, the uniform interleaver which is placed between the two constituent codes in Fig. 4(a) enables one to calculate the distance spectrum and the IOWE of this ensemble, based on the IOWE of the constituent codes (see [3]). The transmission of the codes from this ensemble is assumed to take place over two (independent) parallel binary-input AWGN channels where each bit is equally likely to be assigned to one of these channels ($\alpha_1 = \alpha_2 = \frac{1}{2}$), and the value of the energy per bit to spectral noise density of the first channel is fixed to $\left(\frac{E_b}{N_0}\right)_1 = 0$ dB. Since for long enough block codes, the union bound is not informative at rates beyond the cutoff rate, one would expect that for the considered ensemble of codes (whose block length is roughly 3000 bits), the union bound becomes useless for values of $\left(\frac{E_b}{N_0}\right)_2$ below the value in the RHS of (8) (whose value in this setting is 3.69 dB). This limitation of the union bound is indeed reflected from Fig. 4(b), thus showing how loose is the union bound as compared to computer simulations of the (sub-optimal) iterative decoder. The LMSF bound depicted in Fig. 4(b) uses a partitioning for codes of finite length which was obtained via the algorithm described in Section 5.4; for a bit error probability of $10^{-4}$ it is about 1 dB tighter than the union bound. The DS2 and the 1961 Gallager bounds with their *optimized tilting measures* show a remarkable improvement in their tightness over the union and LMSF bounds where for a bit error probability of $10^{-4}$, the former two bounds exhibit a gain of 0.8 dB over the LMSF bound. The DS2 bound and the 1961 Gallager bound are roughly equally tight;

the DS2 bound gains about 0.05 dB at a bit error probability of $10^{-3}$. In spite of a remarkable advantage of the improved bounds over the union and LMSF bounds, computer simulations under (the sub-optimal) iterative Log-MAP decoding with 10 iterations show a gain of about 0.4 dB, so there is still room for further improvement in the tightness of the bounds under ML decoding.

## 7.2 Distance Properties and Performance Bounds for Various Ensembles of Accumulate-Based Codes

The IOWEs and the distance spectra of ensembles of irregular repeat-accumulate (IRA) codes and accumulate-repeat-accumulate (ARA) codes were derived in [1, 16]. In the continuation of this section, we compare inner bounds on the attainable channel regions of accumulate-based codes under ML decoding. The comparison refers to three ensembles of rate one-third, as depicted in Fig. 5: the first one is the ensemble of uniformly interleaved and non-systematic RA codes where the number of repetitions is $q = 3$, the second and the third ensembles are uniformly interleaved and systematic ensembles of RA codes and ARA codes, respectively, where the number of repetitions is equal to $q = 6$ and, as a result of puncturing, every third bit of the non-systematic part is transmitted (so the puncturing period is $p = 3$). For simplicity of notation, we make use of the abbreviations NSRA$(N, q)$, SPRA$(N, p, q)$ and SPARA$(N, M, p, q)$ codes for the encoders shown in Figs. 5 (a)–(c) (i.e., the abbreviations 'NS' and 'SP' stand for 'non-systematic' and 'systematic and punctured', respectively). In this notation, $N$ is the input block length.

We rely on the concepts of the analysis introduced in [1] for the calculation of the IOWEs of the uniformly interleaved ensembles in Figs. 5 (a)–(c), as well as the calculation of the asymptotic growth rates of their distance spectra. The generalizations of the DS2 and the 1961 Gallager bounds for parallel channels are then applied to these ensembles for the asymptotic case where we let the block length tend to infinity.

The average IOWE and the asymptotic growth rate of the distance spectrum for the ensemble of uniformly interleaved NSRA codes (see Fig. 5 (a)) are given in (16) and (17), respectively.

Following the approach of the analysis in [1], we derive the average IOWEs of the uniformly interleaved ensembles of SPRA$(N,3,6)$ and SPARA$(N,M,3,6)$ codes (see Figs. 5 (b) and (c)). The details of this analysis are provided in Appendix D. In the following, we present the final results related to the finite-length and asymptotic distance properties (where we let $N$ tend to infinity); these results serve later for the calculation of attainable channel regions under ML decoding.

We consider here the case where the repetition code repeats each bit six times, and the parity bits are punctured so that every third bit is transmitted. The IOWE of the ensemble of uniformly interleaved SPRA$(N, 3, 6)$ codes in Fig. 5 (b) is given by (see Appendix D.1)

$$A_{w,d} = \frac{\binom{N}{w}}{\binom{6N}{6w}} \sum_{h=0}^{2N} \sum_{j=0}^{2N} \sum_{i=\max(0,j-2N+h)}^{\min(j,h)} \left\{ \binom{h}{i}\binom{2N-h}{j-i}\binom{2N-d+w}{\lfloor \frac{h}{2} \rfloor} \right. \\ \left. \binom{d-w-1}{\lceil \frac{h}{2} \rceil - 1} 3^{h+j-2i} \, \delta_{6w,2j+h} \right\} \tag{89}$$

where $\delta_{n,m}$ designates the discrete delta function (i.e., it is equal to 1 if $n = m$, and is equal to zero otherwise). The IOWE of the ensemble of uniformly interleaved SPARA$(N, M, 3, 6)$ codes in

Fig. 5 (c) is given by (see Appendix D.1)

$$A_{w,d} = \sum_{m=0}^{M} \sum_{l=0}^{N} \sum_{h=0}^{2N} \sum_{j=0}^{2N} \sum_{i=\max(0,j-2N+h)}^{\min(j,h)} \left\{ \frac{\binom{M}{m}\binom{N-M-l+m}{\lfloor \frac{w-m}{2} \rfloor}\binom{l-m-1}{\lceil \frac{w-m}{2} \rceil - 1}}{\binom{6N}{6l}} \right.$$
$$\left. \binom{h}{i}\binom{2N-h}{j-i}\binom{2N-d+w}{\lfloor \frac{h}{2} \rfloor} \right.$$
$$\left. \binom{d-w-1}{\lceil \frac{h}{2} \rceil - 1} 3^{h+j-2i} \, \delta_{6l,2j+h} \right\}. \qquad (90)$$

The asymptotic growth rates of the distance spectra of these two ensembles are obtained by the calculation of the limit

$$r^{[\mathcal{C}]}(\delta) = \lim_{N \to \infty} \frac{1}{3N} \sum_{w=0}^{N} A_{w,d}, \qquad \delta = \frac{d}{3N} \quad (0 \le \delta \le 1)$$

where it is taken into account that the common rate of these ensembles is equal to one-third bits per channel use.

Let $H(x) = -x \ln(x) - (1-x) \ln(1-x)$ be the binary entropy function to the natural base, then it is well known that

$$\lim_{n \to \infty} \frac{1}{n} \ln \binom{n}{\beta n} = H(\beta), \quad 0 \le \beta \le 1. \qquad (91)$$

For the derivation of the asymptotic growth rate of the average distance spectrum for the ensemble of uniformly interleaved SPRA $(N, 3, 6)$ codes where $N \to \infty$, we rely on the IOWE given in (89), and introduce the parameters

$$\eta \triangleq \frac{h}{3N}, \quad \rho_1 \triangleq \frac{i}{3N}, \quad \rho_2 \triangleq \frac{j}{3N}.$$

The asymptotic growth rate of the distance spectrum of this ensemble is given by (see Appendix D.2)

$$r(\delta) = \max_{\eta, \rho_1, \rho_2} \left\{ -\frac{5}{3} H\left(\frac{2\rho_2 + \eta}{2}\right) + \eta H\left(\frac{\rho_1}{\eta}\right) + \left(\frac{2}{3} - \eta\right) H\left(\frac{\rho_2 - \rho_1}{\frac{2}{3} - \eta}\right) \right.$$
$$+ \left(\frac{2}{3} - \delta + \frac{2\rho_2 + \eta}{6}\right) H\left(\frac{\eta}{2\left(\frac{2}{3} - \delta + \frac{2\rho_2 + \eta}{6}\right)}\right)$$
$$\left. + \left(\delta - \frac{2\rho_2 + \eta}{6}\right) H\left(\frac{\eta}{2\left(\delta - \frac{2\rho_2 + \eta}{6}\right)}\right) + (\eta + \rho_2 - 2\rho_1) \ln 3 \right\} \qquad (92)$$

where the three-parameter maximization is performed over the finite domain which is characterized by the following inequalities:

$$0 \le \eta \le \frac{2}{3}, \quad 0 \le \rho_2 \le \frac{2}{3}, \quad 2\rho_2 + \eta \le 6\delta, \quad \rho_2 + 2\eta \le 3\delta$$
$$\max\left(0, \rho_2 + \eta - \frac{2}{3}\right) \le \rho_1 \le \min(\rho_2, \eta), \quad \eta - \rho_2 + 3\delta \le 2. \qquad (93)$$

Finally, the derivation of the asymptotic growth rate of the average distance spectrum for the ensemble of uniformly interleaved SPARA $(N, M, 3, 6)$ codes, where $N \to \infty$ and the ratio $\frac{M}{N}$ is fixed, relies on the IOWE given in (90). To this end, we introduce the three additional parameters

$$\alpha \triangleq \frac{M}{3N}, \quad \varepsilon_1 \triangleq \frac{m}{3N}, \quad \varepsilon_2 \triangleq \frac{w-m}{3N}. \tag{94}$$

As mentioned above, the value of $\alpha$ is fixed, and also $0 \leq \alpha \leq \frac{1}{3}$ (since $M \leq N$). After straightforward and tedious algebra which is conceptually similar to the calculations in Appendix D.2, one obtains the following expression for the asymptotic growth rate of the average distance spectrum of the considered ensemble of uniformly interleaved SPARA codes:

$$r(\delta) = \max_{\eta, \rho_1, \rho_2, \varepsilon_1, \varepsilon_2} \left\{ \alpha H\left(\frac{\varepsilon_1}{\alpha}\right) + \left(\frac{1}{3} - \alpha - \frac{2\rho_2 + \eta}{6} + \varepsilon_1\right) H\left(\frac{\varepsilon_2}{2\left(\frac{1}{3} - \alpha - \frac{2\rho_2 + \eta}{6} + \varepsilon_1\right)}\right) \right.$$
$$+ \left(\frac{2\rho_2 + \eta}{6} - \varepsilon_1\right) H\left(\frac{\varepsilon_2}{2\left(\frac{2\rho_2 + \eta}{6} - \varepsilon_1\right)}\right) + \eta H\left(\frac{\rho_1}{\eta}\right) - 2H\left(\frac{2\rho_2 + \eta}{2}\right)$$
$$+ \left(\frac{2}{3} - \eta\right) H\left(\frac{\rho_2 - \rho_1}{\frac{2}{3} - \eta}\right) + \left(\frac{2}{3} - \delta + \varepsilon_1 + \varepsilon_2\right) H\left(\frac{\eta}{2\left(\frac{2}{3} - \delta + \varepsilon_1 + \varepsilon_2\right)}\right)$$
$$\left. + (\delta - \varepsilon_1 - \varepsilon_2) H\left(\frac{\eta}{2(\delta - \varepsilon_1 - \varepsilon_2)}\right) + (\eta + \rho_2 - 2\rho_1) \ln 3 \right\} \tag{95}$$

where the five-parameter maximization is performed over the finite domain which is characterized by the following inequalities:

$$0 \leq \eta \leq \frac{2}{3}, \quad 0 \leq \rho_2 \leq \frac{2}{3}, \quad 0 \leq \varepsilon_1 \leq \alpha,$$
$$0 \leq \varepsilon_1 + \varepsilon_2 \leq \min\left(\delta, \frac{1}{3}\right),$$
$$\max\left(0, \rho_2 + \eta - \frac{2}{3}\right) \leq \rho_1 \leq \min(\rho_2, \eta),$$
$$0 \leq \eta \leq \min\left(\frac{4}{3} - 2\delta + 2(\varepsilon_1 + \varepsilon_2), 2\delta - 2(\varepsilon_1 + \varepsilon_2)\right),$$
$$\varepsilon_2 \leq \min\left(\frac{2}{3} - 2\alpha - \frac{2\rho_2 + \eta}{3} + 2\varepsilon_1, \frac{2\rho_2 + \eta}{3} - 2\varepsilon_1\right). \tag{96}$$

In the asymptotic case where we let the block length tend to infinity, inner bounds on the attainable channel regions for the considered ensembles of accumulate-based codes are calculated in this section by Theorem 3.

In Fig. 6, we compare inner bounds on the attainable channel boundaries as calculated by the union, LMSF, 1961 Gallager and DS2 bounds. These plot refers to the ensemble of NSRA$(N, 3)$ codes of rate $\frac{1}{3}$ bits per channel use (see Fig. 5 (a)) where we let $N$ tend to infinity. The asymptotic growth rate of the distance spectrum of this ensemble is calculated by (17) with $q = 3$. The remarkable superiority of the generalized DS2 and 1961 Gallager bounds over the union and LMSF bounds is exemplified for this ensemble of turbo-like codes; actually, the 1961 Gallager bound appears to be slightly tighter than the generalized DS2 bound at the extremities of the boundary of the attainable channel region. This phenomenon was also observed for various turbo-like ensembles, as well as for ensembles of fully random block codes. In the continuation of this section, we therefore compare inner bounds on the attainable channel regions for various ensembles of turbo-like codes where the boundaries of these regions are determined via the generalization of the 1961 Gallager bound (see Fig. 8).

Fig. 7 compares the asymptotic growth rate of the distance spectra of several ensembles of uniformly interleaved and accumulate-based codes where the ensembles are depicted in Fig. 5. The improved performance of the ensembles of SPARA codes under ML decoding is demonstrated by the Gallager bounding technique in Fig. 8. This improvement is attributed to the distance spectral thinning effect [25] which is exemplified in Fig. 7 for the ensembles of NSRA, SPRA and SPARA codes of the same code rate ($\frac{1}{3}$ bits per channel use). The same phenomenon of distance spectral thinning also exists by reducing the value of $\alpha$ for the ensembles of SPARA codes (see Fig. 7, comparing the two plots for $\alpha = \frac{1}{4}$ and $\alpha = \frac{2}{15}$); this in turn yields an improved inner bound on the attainable channel regions, as observed in Fig. 8. It is shown in this figure that for the SPARA ensemble with the parameters $p = 3, q = 6$ and $\alpha = \frac{2}{15}$, the gap between the inner bound on the attainable channel region under ML decoding and the capacity limit is less than 0.05 dB. Note that for the examined ensembles of NSRA and SPRA codes of the same code rate, the corresponding gaps between the inner bounds on the attainable channel regions and the channel capacity are 2.2 dB and 0.5 dB, respectively (see Fig. 8).

## 7.3 Considerations on the Computational Complexity of the Generalized DS2 and 1961 Gallager Bounds

The brute-force calculation of the generalized DS2 bound for linear codes (or ensembles) of finite length is in general computationally heavy. For every constant weight subcode, it requires a numerical optimization over the two parameters $\lambda \geq 0$ and $0 \leq \rho \leq 1$; for each subcode of constant Hamming weight and for each choice of values for $\lambda$ and $\rho$, one needs to solve numerically the explicit equations for $k$ and $\beta_j$ (see Eqs. (43) and (44)) which are related to the $J$ optimized tilting measures. Moreover, for each subcode and a pair of values for $\lambda$ and $\rho$, the evaluation of the generalized DS2 bound requires numerical integrations (or summations, in case the channel outputs are discrete). Performing these tedious and time-consuming optimizations for every constant weight subcode would make the improved bounds less attractive in terms of their practical use for performance evaluation of linear codes and ensembles.

In the following, we suggest an approach which significantly reduces the complexity related to the computation of the generalized DS2 bound, and enhances the applicability of the bound using standard computational facilities. First, the code is partitioned into constant Hamming weight subcodes, and the exact union bound (see Eq. (C.5)) is calculated for every subcode (note that the number of subcodes does not exceed the block length of the code). This task is rather easy, given the (average) distance spectrum $\{A_h\}$ or the weighted IOWE $\{A'_h\}$ of the code (or ensemble) which are calculated in advance (see (23) and (24)). In order to reduce the computational complexity, we do not calculate the generalized DS2 bounds for those constant-Hamming weight subcodes for which the values of the union bounds are below a certain threshold (e.g., we may choose a threshold of $10^{-10}$ for bit error probability or $10^{-6}$ for block error probability; these thresholds should be tailored for the application under consideration). Next, for those constant Hamming weight subcodes for which the union bound exceeds the above threshold, the generalized DS2 bound is evaluated. For these subcodes, we wish to reduce the infinite interval $\lambda \geq 0$ to a finite interval; this is performed by using the transformation $\lambda' \triangleq \frac{\lambda}{\lambda+1}$ so that the two-parameter optimization is reduced to a numerical optimization over the unit square $(\lambda', \rho) \in [0, 1]^2$. In this respect, it was observed that the optimal values of $\lambda'$ and $\rho$ vary rather slowly for consecutive values of the constant Hamming weight $h$, so the search interval associated with the optimization process may be reduced once again with no penalty in the tightness of the bound. In other words, we search for optimal values of $\lambda'$ and $\rho$ only within a neighborhood of the optimal $\lambda'$ and $\rho$ found for the previous subcode. We proceed in this manner until all the relevant subcodes are considered. As an example, we note that for the ensemble of turbo codes depicted in Fig. 4(b), about 80% of the computational time was saved without affecting the numerical results; in this respect, the threshold for the bit error probability analysis was chosen to be $\frac{10^{-6}}{n}$ where $n$ designates the block length of the code. The reduction in the computational complexity becomes however more pronounced for higher SNR values, as the number of subcodes for which the union bound replaces the computation of the generalized DS2 bound increases.

An analogous consideration applies to the generalized version of the 1961 Gallager bound for parallel channels with its related optimized tilting measures.

Referring to the calculation of attainable channel regions, a search over the region of channel parameters is required. As an example, consider a set of parallel AWGN channels characterized by the $J$-tuple of SNRs $(\nu_1, \ldots, \nu_J)$. In order to find the attainable channel boundary, we fix the values of $\nu_1, \ldots, \nu_{J-1}$ and perform a linear search over $\nu_J$ using any appropriate method (e.g., the bisection method) in order to find the smallest value of $\nu_J^*$ for which the lower bound on the error exponent (as obtained by an upper bound on the ML decoding error probability) vanishes. If $(\nu_1, \ldots, \nu_{J-1}, 0)$ is not an attainable point while $(\nu_1, \ldots, \nu_{J-1}, \infty)$ is attainable, then the resulting value $\nu_J^*$ is such that the point $(\nu_1, \ldots, \nu_J^*)$ is on the boundary of the attainable region. The overall complexity of this approach is, of course, polynomial in $J$. We apply this approach in this section for the calculation of inner bounds on the attainable channel regions under ML decoding, referring to the generalizations of the DS2 and 1961 Gallager bounds in Sections 3 and 4, respectively.

## 8 Summary

This paper is focused on the performance analysis of binary linear block codes (or ensembles) whose transmission takes place over independent and memoryless parallel channels. New bounds on the maximum-likelihood (ML) decoding error probability are derived. These bounds are applied to various ensembles of turbo-like codes, focusing especially on repeat-accumulate codes and their recent variations which possess low encoding and decoding complexity and exhibit remarkable performance under iterative decoding (see, e.g., [1, 7, 18, 26]). The framework of the second version of the Duman and Salehi (DS2) bounds is generalized to the case of parallel channels, along with the derivation of their optimized tilting measures. The generalization of the 1961 Gallager bound for parallel channels, introduced by Liu at al. [21], is reviewed and the optimized tilting measures which are related to this bound are calculated via calculus of variations (as opposed to the use of simple and sub-optimal tilting measures in [21]). The connection between the generalized DS2 and the 1961 Gallager bounds, which was originally addressed by Divsalar [8] and by Sason and Shamai [29, 32] for a single channel, is explored for an arbitrary number of independent parallel channels. In this respect, it is shown that the generalization of the DS2 bound for independent parallel channels is not necessarily tighter than the generalization of the 1961 Gallager bound [21], as opposed to the case where the communication takes place over a single channel; this difference is attributed to the additional Jensen's inequality in the transition from (28) to (29) (see p. 13) which is circumvented in the generalization of the 1961 Gallager bound for parallel channels (see (57) in p. 20 where the upper bound is linear in the split weight enumerator, thus there is no need for an additional use of the Jensen's inequality due to the random mapper).

In the asymptotic case where we let the block length tend to infinity, the new bounds are used to obtain improved inner bounds on the attainable channel regions under ML decoding. The tightness of the new bounds for independent parallel channels is exemplified for structured ensembles of turbo-like codes. In this respect, the inner bounds on the attainable channel regions which are computed by the generalized DS2 bound are slightly looser than those computed by the generalized 1961 Gallager bound. On the other hand, for turbo-like ensembles of moderate block lengths, the generalized DS2 bound for parallel channels appears to be slightly tighter than the generalized 1961 Gallager bound (see, e.g., Fig. 4(b) in p. 55). This enhances the conclusion that there is no sharp advantage in the tightness of one bound over the other.

Following the approach in [1], we analyze the distance spectra and their asymptotic growth rates for various ensembles of systematic and punctured accumulate-based codes (see Fig. 5). This distance spectral analysis serves to assess the performance of these codes under ML decoding where we rely on the bounding techniques developed in this paper and in [21] for parallel channels. The improved performance of the ensembles of systematic and punctured accumulate-repeat-accumulate (SPARA) codes under ML decoding is demonstrated by the Gallager bounding technique in Fig. 8. This improvement is attributed to the distance spectral thinning effect [25] which is exemplified in Fig. 7 by comparing the asymptotic growth rates of the distance spectra for the ensembles in Fig. 5 (a)–(c).

The generalization of the DS2 bound for parallel channels enables to re-derive specific bounds which were originally derived by Liu et al. [21] as special cases of the 1961 Gallager bound. The improved bounds with their optimized tilting measures show, regardless of the block length of the codes, an improvement over the union bound and other previously reported bounds for independent parallel channels; this improvement is especially pronounced for moderate to large block lengths. However, in some cases, the new bounds under ML decoding happen to be a bit pessimistic as compared to computer simulations of sub-optimal iterative decoding (see, e.g., Fig. 4(b)), thus indicating that there is still room for further improvement.

# Acknowledgment

This research work was supported by a grant from Intel Israel and by the Taub and Shalom Foundations.

# Appendices

# Appendix A

## A.1 On the Sub-optimality of Even Tilting Measures in the Gallager Bound

In the following, we derive the functions $f(\cdot; j)$ resulting from the optimal DS2 tilting measures in (42) and demonstrate that they are not even functions. From (32), we get the expression

$$\psi(y; j) = \frac{g(y; j)p(y|0; j)}{\sum_{y'} g(y'; j)p(y'|0; j)} \triangleq c^{-1} \cdot g(y; j)p(y|0; j)$$

for the single-letter connection between the normalized and un-normalized DS2 tilting measures; changing the subject gives

$$g(y; j) = c \cdot \left( \frac{\psi(y; j)}{p(y|0; j)} \right) . \tag{A.1}$$

Substituting (42) in (A.1) we obtain the optimal form of the un-normalized tilting measure as

$$g(y; j) = c \cdot \left( 1 + k \left( \frac{p(y|1; j)}{p(y|0; j)} \right)^\lambda \right)^\rho \tag{A.2}$$

Next, we substitute (65) in the LHS of (A.2) and manipulate the expression to get

$$f(y; j) = \text{const} \cdot p(y|0; j) \left( 1 + k \left( \frac{p(y|1; j)}{p(y|0; j)} \right)^\lambda \right)^{\frac{\rho}{s}} . \tag{A.3}$$

Clearly, this expression does not constitute an even function.

## A.2 Some Technical Details Related to the Derivation of the Bound in (67)

In the following, we use the relation between the 1961 Gallager bound and the DS2 bound in order to get a bound similar in form to the DS2 bound (31) which is tighter than (66). We begin with the bounds in (52) and (53) and average their sum over all possible channel assignments, yielding that

$$P_e \leq \mathbf{E} \left\{ \sum_{h=1}^{n} \sum_{\substack{0 \leq h_j \leq n_j \\ \sum h_j = h}} A_{h_1,\ldots,h_J} \prod_{j=1}^{J} [V(r,t;j)]^{h_j} [G(r;j)]^{n_j - h_j} e^{-nrd} + \prod_{j=1}^{J} [G(s;j)]^{n_j} e^{-nsd} \right\}$$

$$= \sum_{\substack{n_j \geq 0 \\ \sum n_j = n}} \left\{ \sum_{h=1}^{n} \sum_{\substack{0 \leq h_j \leq n_j \\ \sum h_j = h}} A_{h_1,\ldots,h_J} \prod_{j=1}^{J} [V(r,t;j)]^{h_j} [G(r;j)]^{n_j - h_j} e^{-nrd} \right.$$

$$\left. + \prod_{j=1}^{J} [G(s;j)]^{n_j} e^{-nsd} \right\} P_{\underline{N}}(\underline{n}) \quad , \quad \begin{array}{c} r \leq 0 \\ s \geq 0 \\ -\infty < d < \infty \end{array} \tag{A.4}$$

where $G(\,\cdot\,;j)$, and $V(\cdot,\cdot\,;j)$ are defined in (64) for $j = 1,\ldots,J$. Following the same steps as in (30)–(31), we get

$$P_{\mathrm{e}} \leq \sum_{h=1}^{n} A_h \left[\sum_{j=1}^{J} \alpha_j V(r,t;j)\right]^h \left[\sum_{j=1}^{J} \alpha_j G(r;j)\right]^{n-h} e^{-nrd}$$
$$+ \left[\sum_{j=1}^{J} \alpha_j G(s;j)\right]^n e^{-nsd}. \qquad (A.5)$$

Optimizing this bound over the parameter $d$ (where $d \in \mathbb{R}$) gives

$$P_{\mathrm{e}} \leq 2^{h(\rho)} \left\{\sum_{h=1}^{n} A_h \left[\sum_{j=1}^{J} \alpha_j V(r,t;j)\right]^h \left[\sum_{j=1}^{J} \alpha_j G(r;j)\right]^{n-h}\right\}^{\rho}$$
$$\cdot \left[\sum_{j=1}^{J} \alpha_j G(s;j)\right]^{n(1-\rho)}. \qquad (A.6)$$

Substituting (65) in (64) yields

$$G(s;j) = \sum_y p(y|0;j) g(y;j)$$
$$G(r;j) = \sum_y p(y|0;j) g(y;j)^{1-\frac{1}{\rho}}$$
$$V(r,t;j) = \sum_y p(y|0;j)^{1-\lambda} p(y|1;j)^{\lambda} g(y;j)^{1-\frac{1}{\rho}}. \qquad (A.7)$$

Finally, substituting (A.7) in (A.6) gives

$$P_{\mathrm{e}} \leq 2^{h(\rho)} \left\{\sum_{h=1}^{n} A_h \left[\sum_{j=1}^{J} \alpha_j \sum_y p(y|0;j)^{1-\lambda} p(y|1;j)^{\lambda} g(y;j)^{1-\frac{1}{\rho}}\right]^h \right.$$
$$\left. \left[\sum_{j=1}^{J} \alpha_j \sum_y p(y|0;j) g(y;j)^{1-\frac{1}{\rho}}\right]^{n-h}\right\}^{\rho}$$
$$\cdot \left[\sum_{j=1}^{J} \alpha_j \sum_y p(y|0;j) g(y;j)\right]^{n(1-\rho)}$$
$$= 2^{h(\rho)} \left\{\sum_{h=0}^{n} \overline{A}_h \left[\sum_{j=1}^{J} \alpha_j \left(\sum_y g(y;j)^{1-\frac{1}{\rho}} p(y|0;j)^{1-\lambda} p(y|1;j)^{\lambda}\right)\right.\right.$$
$$\left(\sum_{j=1}^{J} \alpha_j \sum_y g(y;j) p(y|0;j)\right)^{\frac{1-\rho}{\rho}}\right]^h \left[\sum_{j=1}^{J} \alpha_j \left(\sum_y g(y;j)^{1-\frac{1}{\rho}} p(y|0;j)\right)\right.$$
$$\left.\left.\left(\sum_{j=1}^{J} \alpha_j \sum_y g(y;j) p(y|0;j)\right)^{\frac{1-\rho}{\rho}}\right]^{n-h}\right\}^{\rho}.$$

We note that this bound is tighter than (66), because we have not invoked Jensen's inequality in the process of averaging over all channel assignments in (A.4).

## A.3 Technical Details for Calculus of Variations on (71)

The bound on the decoding error probability for constant Hamming weight codes is given by substituting (70) into (58). Disregarding the multiplicative term $2^{h(\rho)}$, we minimize the expression

$$
\begin{aligned}
U \triangleq A_h & \left\{ \sum_{j=1}^{J} \alpha_j \sum_y [p(y|0;j)p(y|1;j)]^{\frac{1-r}{2}} f(y;j)^r \right\}^h \\
& \cdot \left\{ \sum_{j=1}^{J} \frac{\alpha_j}{2} \sum_y \left[ p(y|0;j)^{1-r} + p(y|1;j)^{1-r} \right] f(y;j)^r \right\}^{n-h} e^{-nrd} \\
& + \left\{ \sum_{j=1}^{J} \frac{\alpha_j}{2} \sum_y \left[ p(y|0;j)^{1-s} + p(y|1;j)^{1-s} \right] f(y;j)^s \right\}^n e^{-nsd},
\end{aligned}
$$
$$ r \leq 0, \; s \geq 0 \; -\infty < d < \infty. \qquad (A.8)$$

Employing calculus of variations, we substitute in (A.8) the following tilting measure

$$ f(y;j) = f_0(y;j) + \varepsilon \eta(y;j) $$

where $\eta(\cdot;j)$ is an arbitrary function. Next, we impose the condition that $\left.\frac{\partial U}{\partial \varepsilon}\right|_{\varepsilon=0} = 0$ for all $\eta(\cdot;j)$. The derivative is given by

$$
\begin{aligned}
\left.\frac{\partial U}{\partial \varepsilon}\right|_{\varepsilon=0} = A_h e^{-nrd} & \left\{ h \left[ \sum_{j=1}^{J} \alpha_j \sum_y (p(y|0;j)p(y|1;j))^{\frac{1-r}{2}} f_0(y;j)^r \right]^{h-1} \right. \\
& \left[ \sum_{j=1}^{J} \alpha_j \sum_y (p(y|0;j)p(y|1;j))^{\frac{1-r}{2}} r f_0(y;j)^{r-1} \eta(y;j) \right] \\
& \left[ \sum_{j=1}^{J} \frac{\alpha_j}{2} \sum_y \left( p(y|0;j)^{1-r} + p(y|1;j)^{1-r} \right) f_0(y;j)^r \right]^{n-h} \\
+ (n-h) & \left[ \sum_{j=1}^{J} \frac{\alpha_j}{2} \sum_y \left( p(y|0;j)^{1-r} + p(y|1;j)^{1-r} \right) f_0(y;j)^r \right]^{n-h-1} \\
& \left[ \sum_{j=1}^{J} \frac{\alpha_j}{2} \sum_y \left( p(y|0;j)^{1-r} + p(y|1;j)^{1-r} \right) r f_0(y;j)^{r-1} \eta(y;j) \right] \\
& \left. \left[ \sum_{j=1}^{J} \alpha_j \sum_y (p(y|0;j)p(y|1;j))^{\frac{1-r}{2}} f_0(y;j)^r \right]^h \right\} \\
+ e^{-nsd} n & \left[ \sum_{j=1}^{J} \frac{\alpha_j}{2} \sum_y \left( p(y|0;j)^{1-s} + p(y|1;j)^{1-s} \right) f_0(y;j)^s \right]^{n-1} \\
& \left[ \sum_{j=1}^{J} \frac{\alpha_j}{2} \sum_y \left( p(y|0;j)^{1-s} + p(y|1;j)^{1-s} \right) s f_0(y;j)^{s-1} \eta(y;j) \right]. \qquad (A.9)
\end{aligned}
$$

Defining the constants

$$c_1 \triangleq A_h e^{-nrd} hr \left[ \sum_{j=1}^J \alpha_j \sum_y (p(y|0;j)p(y|1;j))^{\frac{1-r}{2}} f_0(y;j)^r \right]^{h-1}$$

$$c_2 \triangleq \left[ \sum_{j=1}^J \frac{\alpha_j}{2} \sum_y \left( p(y|0;j)^{1-r} + p(y|1;j)^{1-r} \right) f_0(y;j)^r \right]^{n-h}$$

$$c_3 \triangleq A_h e^{-nrd} \frac{r(n-h)}{2} \left[ \sum_{j=1}^J \frac{\alpha_j}{2} \sum_y \left( p(y|0;j)^{1-r} + p(y|1;j)^{1-r} \right) f_0(y;j)^r \right]^{n-h-1}$$

$$c_4 \triangleq \left[ \sum_{j=1}^J \alpha_j \sum_y (p(y|0;j)p(y|1;j))^{\frac{1-r}{2}} f_0(y;j)^r \right]^{h}$$

$$c_5 \triangleq e^{-nsd} \frac{ns}{2} \left[ \sum_{j=1}^J \frac{\alpha_j}{2} \sum_y \left( p(y|0;j)^{1-s} + p(y|1;j)^{1-s} \right) f_0(y;j)^s \right]^{n-1}$$

(A.10)

and requiring that the integrand in (A.9) be equal to zero, we get the equivalent condition

$$\sum_{j=1}^J \alpha_j \left\{ \left( c_1 c_2 \left[ p(y|0;j)p(y|1;j) \right]^{\frac{1-r}{2}} + c_3 c_4 \left[ p(y|0;j)^{1-r} + p(y|1;j)^{1-r} \right] \right) f_0(y;j)^{r-1} \right.$$

$$\left. + c_5 \left[ p(y|0;j)^{1-s} + p(y|1;j)^{1-s} \right] f_0(y;j)^{s-1} \right\} = 0, \quad \forall y \in \mathcal{Y}.$$

Defining $K_1 \triangleq \frac{c_1 c_2}{c_5}$, $K_2 \triangleq \frac{c_3 c_4}{c_5}$, and dividing both sides by $f_0(y;j)^{r-1}$ implies the condition in (72).

# Appendix B

### Proof of Theorem 3

The concept of the proof of this theorem is similar to the proofs introduced in [20, pp. 40–42] for the single channel case, and the proofs of [21, Theorems 2–4] for the scenario of independent parallel channels. The difference in this proof from those mentioned above is the starting point which relies on the generalization of the DS2 bound (see Theorem 1 in Section 3). Like these proofs, the essence of the proof is showing that apart from the requirement in (86), the lower bound on the error exponent in (37), which follows here from the generalized DS2 bound, is positive and behaves linearly in $\delta$ for small enough values of $\delta$. To this end, let us first verify from (43) that the partial derivative of the optimized parameter $k$ w.r.t. $\delta$ is strictly positive at $\delta = 0$. This follows by rewriting the implicit equation for $k$ in (43) as

$$k = \frac{\delta}{1-\delta} \frac{h(k)}{g(k)}$$

where $h$ and $g$ stand for the numerator and denominator of the term which multiplies $\frac{\delta}{1-\delta}$ in the RHS of (43). Note that (43) also implies that $k = 0$ at $\delta = 0$, and hence at $\delta = 0$, (44) gives that $\beta_j = 1$ for all $j \in \{1, \ldots, J\}$. Differentiating both sides of (43) w.r.t. $\delta$ and setting $\delta = 0$ gives

$$\left.\frac{\partial k}{\partial \delta}\right|_{\delta=0} = \left\{\sum_{j=1}^{J} \sum_{y \in \mathcal{Y}} \alpha_j p(y|0;j)^{1-\lambda} p(y|1;j)^{\lambda}\right\}^{-1} \triangleq \eta. \tag{B.1}$$

From (87), it follows that $\limsup_{\delta \to 0} r^{[\mathcal{C}]}(\delta) \leq 0$, so in the limit where $\delta \to 0$, the Bhattacharyya union bound becomes tight for the conditional ML decoding error probability (given that the all-zero codeword is transmitted) w.r.t. subcodes of constant Hamming weight with normalized Hamming weight $\delta$. Hence, $\rho \to 1$ and $\lambda \to \frac{1}{2}$ become optimal in the limit where $\delta \to 0$. The substitution of $\lambda = \frac{1}{2}$ in the RHS of (B.1) gives

$$\left.\frac{\partial k}{\partial \delta}\right|_{\delta=0} = \left\{\sum_{j=1}^{J} \alpha_j \gamma_j\right\}^{-1}$$

where $\gamma_j$ is the Bhattacharyya parameter of the $j$-th channel.

For values of $\delta$ close enough to zero, let us analyze the behavior of each of the three terms in the RHS of (37). Let

$$r_0 \triangleq \limsup_{\delta \to 0} \frac{r^{[\mathcal{C}]}(\delta)}{\delta} \tag{B.2}$$

then the first term in the RHS of (37) behaves like $-r_0 \rho \delta$ for values of $\delta$ close enough to zero, i.e.,

$$-\rho r^{[\mathcal{C}]}(\delta) = -r_0 \rho \delta + O(\delta^2). \tag{B.3}$$

As for the second term in the RHS of (37), since $k$ tends to zero and $\beta_j \to 1$ for all $j$ as we let $\delta \to 0$, then the optimized tilting measures $\psi(y;j)$ in (42) tend to the conditional pdfs $p(y|0;j)$, respectively, for all $j \in \{1, \ldots, J\}$. Since $\rho \to 1$ and $\lambda \to \frac{1}{2}$ are optimal as we let $\delta \to 0$, then in the limit where $\delta$ is close enough to zero, we get

$$-\rho \delta \ln \left(\sum_{j=1}^{J} \alpha_j \sum_{y \in \mathcal{Y}} \psi(y;j)^{1-\frac{1}{\rho}} p(y|0;j)^{\frac{1-\lambda\rho}{\rho}} p(y|1;j)^{\lambda}\right)$$

$$= -\rho \delta \ln \left(\sum_{j=1}^{J} \alpha_j \sum_{y \in \mathcal{Y}} \sqrt{p(y|0;j) p(y|1;j)}\right) + o(\delta)$$

$$= -\rho \delta \ln \left(\sum_{j=1}^{J} \alpha_j \gamma_j\right) + o(\delta). \tag{B.4}$$

As we let $\delta$ tend to zero, we will show that the third term in the RHS of (37) converges to zero quadratically in $\delta$. To this end, we rely on the expression for the optimized tilting measure in (42) and the linear behavior of $k$ near $\delta$ (from (B.1) and since $k = 0$ for $\delta = 0$, it follows that $k = \eta \delta + O(\delta^2)$). We obtain from (42) that for values of $\delta$ close enough to zero, the optimized tilting measure of the generalized DS2 bound behaves like

$$\psi(y;j) = \beta_j p(y|0;j) \left(1 + \eta \delta \sqrt{\frac{p(y|1;j)}{p(y|0;j)}}\right) + O(\delta^2) \tag{B.5}$$

where $\beta_j$ is the normalization factor which is calculated by (44) and is given by $\beta_j = \frac{1}{1+\eta\delta\gamma_j}$. Hence, for small values of $\delta$ which are close enough to zero, the third term in the RHS of (37) behaves like

$$-\rho(1-\delta)\ln\left(\sum_{j=1}^{J}\alpha_j\sum_{y\in\mathcal{Y}}\psi(y;j)^{1-\frac{1}{\rho}}p(y|0;j)^{\frac{1}{\rho}}\right)$$

$$\stackrel{(a)}{=} -\rho(1-\delta)\ln\left(\sum_{j=1}^{J}\alpha_j\sum_{y\in\mathcal{Y}}(1+\eta\delta\gamma_j)^{\frac{1}{\rho}-1}p(y|0;j)\left(1+\eta\delta\sqrt{\frac{p(y|1;j)}{p(y|0;j)}}\right)^{1-\frac{1}{\rho}}\right) + O(\delta^2)$$

$$\stackrel{(b)}{=} -\rho(1-\delta)\ln\left(\sum_{j=1}^{J}\alpha_j\left(1+\eta\left(\frac{1}{\rho}-1\right)\delta\gamma_j\right)\sum_{y\in\mathcal{Y}}\left\{p(y|0;j)\left(1+\eta\delta\left(1-\frac{1}{\rho}\right)\sqrt{\frac{p(y|1;j)}{p(y|0;j)}}\right)\right\}\right) + O(\delta^2)$$

$$\stackrel{(c)}{=} -\rho(1-\delta)\ln\left(\sum_{j=1}^{J}\alpha_j\left(1+\eta\left(\frac{1}{\rho}-1\right)\gamma_j\delta\right)\left(1-\eta\left(\frac{1}{\rho}-1\right)\gamma_j\delta\right)\right) + O(\delta^2)$$

$$= -\rho(1-\delta)\ln\left(\sum_{j=1}^{J}\alpha_j\left(1-\eta^2\left(\frac{1}{\rho}-1\right)^2\gamma_j^2\delta^2\right)\right) + O(\delta^2)$$

$$= -\rho(1-\delta)\ln\left(1-\eta^2\left(\frac{1}{\rho}-1\right)^2\sum_{j=1}^{J}\alpha_j\gamma_j^2\,\delta^2\right) + O(\delta^2)$$

$$\stackrel{(d)}{=} \rho(1-\delta)\eta^2\left(\frac{1}{\rho}-1\right)^2\sum_{j=1}^{J}\alpha_j\gamma_j^2\,\delta^2 + O(\delta^2) = O(\delta^2) \tag{B.6}$$

where (a) follows from (B.5), (b) follows from the equality

$$(1+\epsilon\delta)^\zeta = 1 + \epsilon\zeta\delta + O(\delta^2), \quad \forall \epsilon, \zeta \in \mathbb{R}$$

(c) follows from the definition of the Bhattacharyya constant $\gamma_j$ for the $j$-th channel and since $\sum_{y\in\mathcal{Y}}p(y|0;j) = 1$, and (d) follows from the equality $\ln(1+x) = x + O(x^2)$. Hence, as we let $\delta \to 0$, the expression in (B.6) converges to zero quadratically in $\delta$.

From (B.3)–(B.6), one obtains that for values of $\delta$ which are close enough to zero, the error exponent in (37) decays linearly with $\delta$ if and only if

$$-r_0 - \ln\left(\sum_{j=1}^{J}\alpha_j\gamma_j\right) > 0$$

which coincides with the condition in (87) ($r_0$ is defined in (B.2)). This is indeed the second requirement in Theorem 3. By combining it with the requirement in (86), we obtain that the ML decoding error probability vanishes as we let $n \to \infty$, as long as we exclude the codewords whose Hamming weights behave sub-linearly with the block length $n$. By showing that the error exponent of the generalized DS2 bound grows linearly with $\delta$ as we let $\delta \to 0$ and due to the requirement in (88), it follows that asymptotically in the limit where the block length tends to infinity, these codewords contribute a vanishing effect to the ML decoding error probability (similarly to the proofs of [20, Theorem 2.3] and [21, Theorems 2–4]).

# Appendix C

## Exact Union Bound for Parallel Gaussian Channels

In this appendix, we derive the union bound on the ML decoding error probability of binary linear block codes transmitted over parallel Gaussian channels. This form of the union bound will be used in conjunction with other bounds (e.g., 1961 Gallager or the DS2 bounds) for constant Hamming weight subcodes in order to tighten the resulting bound.

We start the derivation by expressing the pairwise error probability given that the all-zero codeword is transmitted

$$P_e(\underline{0} \to \underline{x}_{h_1,h_2,\ldots,h_J}) = Q\left(\sqrt{2\sum_{j=1}^{J} \nu_j h_j}\right) \tag{C.1}$$

where $\underline{x}_{h_1,h_2,\ldots,h_J}$ is a codeword possessing split Hamming weights $h_1,\ldots,h_J$ in the $J$ parallel channels, and $\nu_j \triangleq \left(\frac{E_s}{N_0}\right)_j$ designates the energy per symbol to spectral noise density for the $j^{\text{th}}$ AWGN channel ($j = 1, 2, \ldots, J$). The union bound on the block error probability gives

$$P_e \leq \sum_{h=1}^{n} \sum_{\substack{h_1 \geq 0, \ldots, h_J \geq 0 \\ h_1 + \ldots + h_J = h}} A_{h_1,\ldots,h_J} Q\left(\sqrt{2\sum_{j=1}^{J} \nu_j h_j}\right) \tag{C.2}$$

where this bound is expressed in terms of the split weight enumerator of the code. Averaging (C.2) over all possible channel assignments and codes from the ensemble gives (see (30))

$$\begin{aligned}
\overline{P_e} &\leq \sum_{\substack{n_j \geq 0 \\ n_1 + \ldots + n_J = n}} \left\{ \sum_{h=1}^{n} \sum_{\substack{0 \leq h_j \leq n_j \\ \sum_j h_j = h}} \overline{A_h}\, P_{\underline{H}|\underline{N}}(\underline{h}|\underline{n})\, P_{\underline{N}}(\underline{n})\, Q\left(\sqrt{2\sum_{j=1}^{J} \nu_j h_j}\right) \right\} \\
&= \sum_{\substack{n_j \geq 0 \\ n_1 + \ldots + n_J = n}} \left\{ \sum_{h=1}^{n} \sum_{\substack{0 \leq h_j \leq n_j \\ \sum_j h_j = h}} \overline{A_h} \binom{h}{h_1, h_2, \ldots, h_J} \right. \\
&\qquad\qquad \binom{n-h}{n_1 - h_1, n_2 - h_2, \ldots, n_J - h_J} \\
&\qquad\qquad \left. \alpha_1^{n_1} \ldots \alpha_J^{n_J} Q\left(\sqrt{2\sum_{j=1}^{J} \nu_j h_j}\right) \right\}
\end{aligned} \tag{C.3}$$

where $\alpha_j$ designates the a-priori probability for the transmission of symbols over the $j^{\text{th}}$ channel, assuming the assignments of these symbols to the $J$ parallel channels are independent and random.

In order to simplify the final result, we rely on Craig's identity for the $Q$-function, i.e.,

$$Q(x) = \frac{1}{\pi} \int_0^{\frac{\pi}{2}} e^{-\frac{x^2}{2\sin^2\theta}} d\theta, \qquad x \geq 0. \tag{C.4}$$

Plugging (C.4) into (C.3) and interchanging the order of integration and summation gives

$$
\begin{aligned}
\overline{P_{\text{e}}} &\leq \frac{1}{\pi} \int_0^{\frac{\pi}{2}} \sum_{\substack{n_j \geq 0 \\ n_1 + \ldots + n_J = n}} \left\{ \sum_{h=1}^{n} \sum_{\substack{0 \leq h_j \leq n_j \\ \sum h_j = h}} \overline{A_h} \binom{h}{h_1, h_2, \ldots, h_J} \right. \\
&\qquad \left. \binom{n-h}{n_1 - h_1, n_2 - h_2, \ldots, n_J - h_J} \alpha_1^{n_1} \cdots \alpha_J^{n_J} \prod_{j=1}^{J} e^{-\frac{\nu_j h_j}{\sin^2 \theta}} \right\} \, d\theta \\
&\stackrel{(a)}{=} \frac{1}{\pi} \int_0^{\frac{\pi}{2}} \sum_{h=1}^{n} \overline{A_h} \sum_{\substack{h_j \geq 0 \\ \sum h_j = h}} \left\{ \binom{h}{h_1, h_2, \ldots, h_J} \prod_{j=1}^{J} \left[ \alpha_j e^{-\frac{\nu_j}{\sin^2 \theta}} \right]^{h_j} \right\} \\
&\qquad \sum_{\substack{k_j \geq 0 \\ \sum_j k_j = n - h}} \left\{ \binom{n-h}{k_1, k_2, \ldots, k_J} \prod_{j=1}^{J} (\alpha_j)^{k_j} \right\} d\theta \\
&\stackrel{(b)}{=} \frac{1}{\pi} \int_0^{\frac{\pi}{2}} \sum_{h=1}^{n} \overline{A_h} \left[ \sum_{j=1}^{J} \alpha_j e^{-\frac{\nu_j}{\sin^2 \theta}} \right]^{h} d\theta \qquad (C.5)
\end{aligned}
$$

where (a) follows by substituting $k_j = n_j - h_j$ for $j = 1, 2, \ldots, J$, and (b) follows since the sequence $\{\alpha_j\}_{j=1}^{J}$ is a probability distribution, which gives the equality

$$
\sum_{\substack{k_j \geq 0 \\ \sum_j k_j = n - h}} \left\{ \binom{n-h}{k_1, k_2, \ldots, k_J} \prod_{j=1}^{J} (\alpha_j)^{k_j} \right\} = \left( \sum_{j=1}^{J} \alpha_j \right)^{n-h} = 1.
$$

Eq. (C.5) provides the exact ($Q$-form) version of the union bound on the block error probability for independent parallel AWGN channels.

# Appendix D

## Distance Spectra Analysis of Systematic Accumulate-Based Codes with Puncturing

The following analysis is focused on the distance spectra of uniformly interleaved and systematic ensembles of repeat-accumulate (RA) codes and accumulate-repeat-accumulate (ARA) codes with puncturing (see Figs. 5 (b) and (c) in p. 56). As mentioned in Section 7.2, these two ensembles are abbreviated by SPRA and SPARA codes, respectively (where 'SP' stands for 'systematic and punctured'). We derive here the input-output weight enumerator (IOWEs) of these ensembles and also calculate the asymptotic growth rates of their distance spectra. The analysis follows the approach introduced in [1], and it is written in a self-contained manner.

The component codes constructing SPRA and SPARA codes are an accumulate code (i.e., a rate-1 differential encoder), a repetition code and a single parity-check (SPC) code. Since we consider ensembles of uniformly interleaved codes, their IOWEs depend on the IOWE of the above component codes [3, 4]. As a preparatory step, we introduce the IOWEs of these components.

1. The IOWE of a repetition (REP) code is given by

$$
A_{w,d}^{\text{REP}(q)} = \binom{k}{w} \delta_{d,qw} \qquad (D.1)
$$

where $k$ designates the input block length, and $\delta_{n,m}$ is the discrete delta function.

2. The IOWE of an accumulate (ACC) code is given by

$$A_{w,d}^{\text{ACC}} = \binom{n-d}{\lfloor \frac{w}{2} \rfloor}\binom{d-1}{\lceil \frac{w}{2} \rceil - 1} \tag{D.2}$$

where $n$ is the block length (since this code is of rate 1, the input and output block lengths are the same). The IOWE in (D.2) can be easily obtained combinatorially; to this end, we rely on the fact that for the accumulate code, every single '1' at the input sequence flips the value at the output from this point (until the occurrence of the next '1' at the input sequence).

3. The IOWE function of a non-systematic single parity-check code which provides the parity bit of each set of $p$ consecutive bits, call it $\text{SPC}(p)$, is given by (see [1, Eq. (8)])

$$\begin{aligned} A(W,D) &= \sum_{w=0}^{np}\sum_{d=0}^{n} A_{w,d}^{\text{SPC}(p)} W^w D^d \\ &= \left[\text{Even}\big((1+W)^p\big) + \text{Odd}\big((1+W)^p\big)D\right]^n \end{aligned} \tag{D.3}$$

where

$$\begin{aligned} \text{Even}\big((1+W)^p\big) &= \frac{(1+W)^p + (1-W)^p}{2} \\ \text{Odd}\big((1+W)^p\big) &= \frac{(1+W)^p - (1-W)^p}{2} \end{aligned} \tag{D.4}$$

are two polynomials which include the terms with the even and odd powers of $W$, respectively.

To verify (D.3), note that a parity-bit of this code is equal to 1 if and only if the number of ones in the corresponding set of $p$ bits is odd; also, the number of check nodes in the considered code is equal to the block length of the code ($n$).

The case where the output bits of an accumulate code are punctured with a puncturing period $p$ is equivalent to an $\text{SPC}(p)$ code followed by an accumulate code (see Fig. 9 which was originally shown in [1, Fig. 2]). Hence, for the uniformly interleaved ensembles of SPRA and SPARA codes with a puncturing period of $p=3$ (see Figs. 5 (b) and (c)), we are interested in the IOWE of the $\text{SPC}(3)$ code. For the case where $p=3$, (D.4) gives

$$\text{Even}\big((1+W)^3\big) = 1 + 3W^2, \quad \text{Odd}\big((1+W)^3\big) = 3W + W^3$$

and (D.3) gives straightforwardly the following IOWE of the $\text{SPC}(3)$ code [1, Eq. (15)]:

$$A_{w,d}^{\text{SPC}(3)} = \binom{n}{d}\sum_{j=0}^{n}\sum_{i=\max(0,j-n+d)}^{\min(j,d)} \binom{d}{i}\binom{n-d}{j-i} 3^{d+j-2i}\, \delta_{w,2j+d}. \tag{D.5}$$

In the following, we consider two uniformly interleaved ensembles of SPRA and SPARA codes with $q=6$ repetitions and puncturing of period $p=3$, as shown in Fig. 5 (b) and (c). We rely here on the equivalence shown in Fig. 9, related to the inner accumulate code with puncturing. In this respect, since the input bits to the SPC (appearing in the right plot in Fig. 9) are permuted by the uniform interleaver which is placed after the repetition code (see Figs. 5 (b) and (c)), then the average IOWEs of these two ensembles remain unaffected by placing an additional uniform interleaver between the SPC and the inner accumulate codes. Similarly, placing another uniform interleaver between the precoder in Fig. 5 (c) (i.e., the code which accumulates the first $N-M$ bits) and the repetition code, does not affect the average IOWE of the overall ensemble in Fig. 5 (c).

As mentioned above, the equivalence in Fig. 9 yields that without loss of generality, an additional uniform interleaver of length $N' = \frac{qN}{p} = 2N$ bits can be placed between the SPC(3) code and the accumulate code without affecting the calculation. By doing so, the average IOWE of the serially concatenated and uniformly interleaved ensemble whose constituent codes are the SPC(3) and the accumulate codes, call it ACC(3), is given by (see [4])

$$A_{w,d}^{\text{ACC}(3)} = \sum_{h=0}^{2N} \frac{A_{w,h}^{\text{SPC}(3)} A_{h,d}^{\text{ACC}}}{\binom{2N}{h}}. \tag{D.6}$$

The substitution of (D.2) and (D.5) into (D.6) gives

$$A_{w,d}^{\text{ACC}(3)} = \sum_{h=0}^{2N} \sum_{j=0}^{2N} \sum_{i=\max(0,j-2N+h)}^{\min(j,h)} \left\{ \binom{h}{i} \binom{2N-h}{j-i} \binom{2N-d}{\lfloor \frac{h}{2} \rfloor} \binom{d-1}{\lceil \frac{h}{2} \rceil - 1} \right. \\ \left. 3^{h+j-2i} \, \delta_{w,2j+h} \right\}. \tag{D.7}$$

Note that (D.7) is similar to [1, Eq. (19)], except that $N$ in the latter equation is replaced by $2N$ in (D.7). This follows since $\frac{q}{p}$ (i.e., the ratio between the number of repetitions and the puncturing period) is equal here to 2, instead of 1 as was the case in [1]. Equation (D.7) will be used for the finite-length analysis of the distance spectra for the ensembles considered in the continuation to this appendix.

## D.1 Finite-Length Analysis of the Distance Spectra for Systematic Ensembles of RA and ARA Codes with Puncturing

*Uniformly Interleaved SPRA $(N, 3, 6)$ codes*: Let us consider the ensemble depicted in Fig. 5 (b) where $q = 6$ and $p = 3$. Since there is a uniform interleaver of length $N'' = qN$ between the repetition code and the equivalent ACC(3) code, the average IOWE of this serially concatenated and uniformly interleaved systematic ensemble is given by

$$\begin{aligned} A_{w,d}^{\text{SPRA}(N,3,6)} &= \sum_{l=0}^{6N} \frac{A_{w,l}^{\text{REP}(6)} A_{l,d-w}^{\text{ACC}(3)}}{\binom{6N}{l}} \\ &= \frac{\binom{N}{w} A_{6w,d-w}^{\text{ACC}(3)}}{\binom{6N}{6w}} \end{aligned} \tag{D.8}$$

where the last equality is due to the equality in (D.1). Substituting (D.7) in the RHS of (D.8) gives the average IOWE of the considered ensemble, and this result coincides with (89).

*Uniformly Interleaved SPARA $(N, M, 3, 6)$ codes*: By comparing Figs. 5 (b) and (c), we see that a precoder is placed in the second figure. Referring to the ensemble of SPARA codes which is shown in Fig. 5 (c), the precoder is a binary linear block code whose first $N - M$ input bits are accumulated and the other $M$ input bits not changed; these $N$ bits are encoded by the repetition code. The IOWE of the systematic precoder, call it Pre$(N,M)$, is given by

$$A_{w,d}^{\text{Pre}(N,M)} = \sum_{m=0}^{M} \binom{M}{m} A_{w-m,d-m}^{\text{ACC}}$$
$$= \sum_{m=0}^{M} \left\{ \binom{M}{m} \binom{N-M-d+m}{\lfloor \frac{w-m}{2} \rfloor} \binom{d-m-1}{\lceil \frac{w-m}{2} \rceil - 1} \right\} \quad (D.9)$$

where the last equality relies on (D.2). As mentioned before, for the uniformly interleaved SPARA ensemble depicted in Fig. 5 (c), an additional uniform interleaver between the precoder and the following stages of its encoder does not affect the average IOWE; this ensemble can be therefore viewed as a serial concatenation with a uniform interleaver of length $N$ which is placed between the precoder and the repetition code in Fig. 5 (c) (in addition to the uniform interleaver which is placed after the repetition code). Moreover, referring to the systematic ensemble whose components are REP(6) and ACC(3), the input bits (which are the bits provided by the precoder to the second stage in Fig. 5 (c)) are not transmitted to the channel. In light of these two observations, the average IOWE of the uniformly interleaved ensemble of SPARA codes shown in Fig. 5 (c) is given by

$$A_{w,d}^{\text{SPARA}(N,M,3,6)} = \sum_{l=0}^{N} \frac{A_{w,l}^{\text{Pre}(N,M)} A_{l,d-w+l}^{\text{SPRA}(N,3,6)}}{\binom{N}{l}}. \quad (D.10)$$

By substituting (D.8) (i.e., the equality in (89)) and (D.9) into (D.10), one obtains the expression in (90) for the average IOWE of the SPARA$(N,M,3,6)$ codes.

## D.2 Asymptotic Analysis of the Distance Spectra

This subsection considers the calculation of the asymptotic growth rates of the distance spectra for the two ensembles in Figs. 5 (b) and (c). The calculation of the asymptotic growth rate of the distance spectrum of a sequence of codes (or ensembles) is performed via (14). In the following, we exemplify the derivation of (92) from the average IOWE in (89). The derivation of (95) from (90) is conceptually similar, but is more tedious algebraically. Since we focus here on ensembles of rate one-third and the block length of the input bits is $N$ (see Fig. 5), the asymptotic growth rate of their distance spectra is obtained by normalizing the logarithm of the average distance spectra of the considered ensemble by $n = 3N$ and letting $N$ tend to infinity. Referring to the average IOWE of the uniformly interleaved ensemble of SPRA$(N,3,6)$ codes, as given in (89), we introduce the normalized parameters

$$\delta \triangleq \frac{d}{3N}, \quad \eta \triangleq \frac{h}{3N}, \quad \rho_1 \triangleq \frac{i}{3N}, \quad \rho_2 \triangleq \frac{j}{3N}. \quad (D.11)$$

The normalization by $3N$ yields that the new parameters satisfy

$$0 \leq \delta \leq 1, \quad 0 \leq \eta \leq \frac{2}{3}, \quad 0 \leq \rho_2 \leq \frac{2}{3}. \quad (D.12)$$

From the partial sum w.r.t. the index $i$ in the RHS of (89), dividing the terms in the inequality

$$\max(0, j - 2N + h) \leq i \leq \min(j, h)$$

by $3N$ gives

$$\max\left(0, \rho_2 + \eta - \frac{2}{3}\right) \leq \rho_1 \leq \min(\rho_2, \eta). \quad (D.13)$$

Since the codes are systematic and the block length of the input bits is $N$, we get that the terms which contribute to the IOWE in the RHS of (89) satisfy

$$w \leq \min(d, N), \quad 6w = 2j + h \tag{D.14}$$

and, from (D.11), multiplying (D.14) by $\frac{1}{3N}$ gives

$$\frac{2\rho_2 + \eta}{6} \leq \min\left(\delta, \frac{1}{3}\right). \tag{D.15}$$

From the binomial coefficients which appear in the RHS of (89), it is required that

$$2N - d + w \geq \left\lfloor \frac{h}{2} \right\rfloor, \quad d - w \geq \left\lceil \frac{h}{2} \right\rceil$$

so dividing both sides of these inequalities by $3N$, and letting $N$ tend to infinity gives

$$\eta - \rho_2 + 3\delta \leq 2, \quad \rho_2 + 2\eta \leq 3\delta. \tag{D.16}$$

Combining (D.12)–(D.16) gives the domain for the three parameters $\eta$, $\rho_1$ and $\rho_2$ in (93).

A marginalization of the IOWE enables one to obtain the distance spectrum

$$A_d = \sum_{w=0}^{N} A_{w,d} \tag{D.17}$$

where the IOWE $\{A_{w,d}\}$ is given by (89). Note that unless

$$\frac{w}{N} = \frac{2j + h}{6N} = \frac{2\rho_2 + \eta}{2} \tag{D.18}$$

the IOWE $A_{w,d}$ in (89) vanishes, and therefore it does not affect the sum in the RHS of (D.17). In the limit where $N \to \infty$, the asymptotic growth rate of the average distance spectrum for the uniformly interleaved ensemble of SPRA$(N, 3, 6)$ codes (see Fig. 5 (b) in p. 56) is obtained from (89), (91), and (D.17). Hence, we get

$$\begin{aligned}
r(\delta) &= \lim_{N \to \infty} \frac{1}{3N} \ln \sum_{w=0}^{N} A_{w,d} \\
&= \lim_{N \to \infty} \max_{h,i,j} \Bigg\{ \frac{1}{3N} \bigg[ NH\left(\frac{w}{N}\right) - 6NH\left(\frac{6w}{6N}\right) + hH\left(\frac{i}{h}\right) + (2N - h)H\left(\frac{j-i}{2N-h}\right) \\
&\quad + (2N - d + w)H\left(\frac{h}{2(2N - d + w)}\right) + (d - w - 1)H\left(\frac{\frac{h}{2} - 1}{d - w - 1}\right) \\
&\quad + (h + j - 2i) \ln 3 \bigg] \Bigg\}.
\end{aligned}$$

By multiplying the three parameters which are involved in the maximization by $\frac{1}{3N}$, using (D.11) and (D.18) and taking the limit where $N$ tends to infinity, one readily obtains the result in (92).

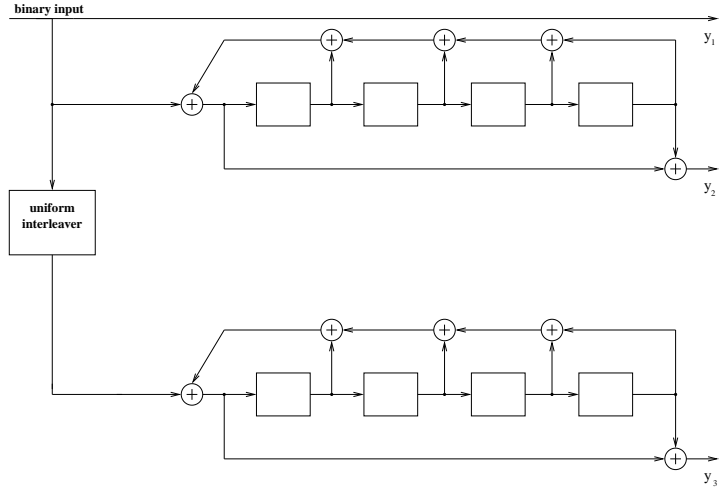

(a) Ensemble of uniformly interleaved turbo codes.

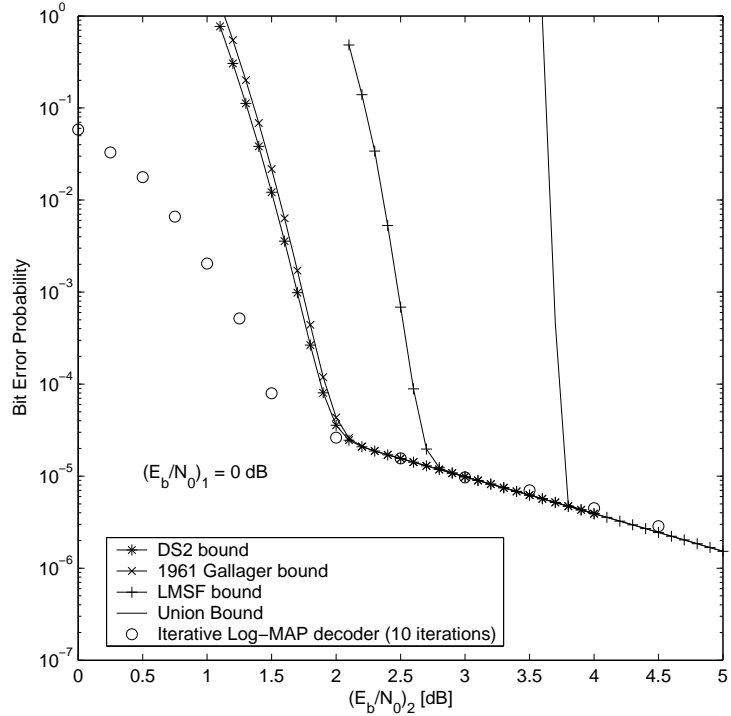

(b) Performance Bounds under ML decoding versus simulation results of iterative Log-MAP decoding.

Figure 4: (a) The encoder of an ensemble of uniformly interleaved turbo codes whose interleaver is of length 1000, and there is no puncturing of parity bits. (b) Performance bounds for the bit error probability under ML decoding versus computer simulation results of iterative Log-MAP decoding (with 10 iterations). The transmission of this ensemble takes place over two (independent) parallel binary-input AWGN channels. Each bit is equally likely to be assigned to one of these channels, and the energy per bit to spectral noise density of the first channel is set to $\left(\frac{E_b}{N_0}\right)_1 = 0$ dB. The compared upper bounds on the bit error probability are the generalizations of the DS2 and 1961 Gallager bounds, the LMSF bound from [21], and the union bound (based on (C.5)).

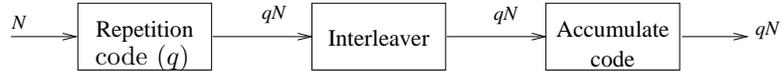

(a) Non-systematic RA codes - NSRA $(N, q)$

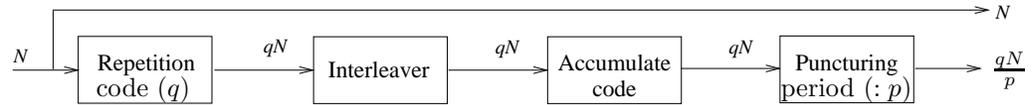

(b) Systematic RA codes with puncturing - SPRA $(N, p, q)$

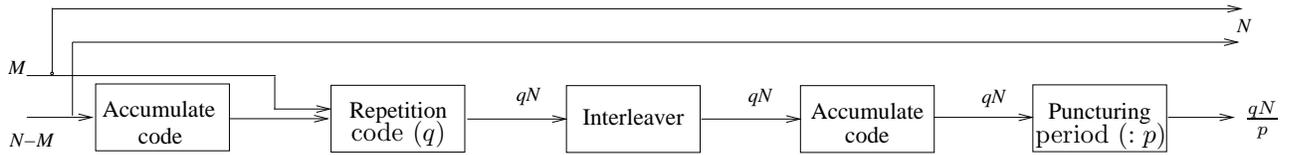

(c) Systematic ARA codes with puncturing - SPARA $(N, M, p, q)$

Figure 5: Systematic and Non-systematic RA and ARA codes. The interleavers of these ensembles are assumed to be chosen uniformly at random, and are of length $qN$ where $N$ designates the length of the input block (information bits) and $q$ is the number of repetitions. The rates of all the ensembles is set to $\frac{1}{3}$ bits per channel use, so we set $q = 3$ for figure (a), and $q = 6$ and $p = 3$ for figures (b) and (c).

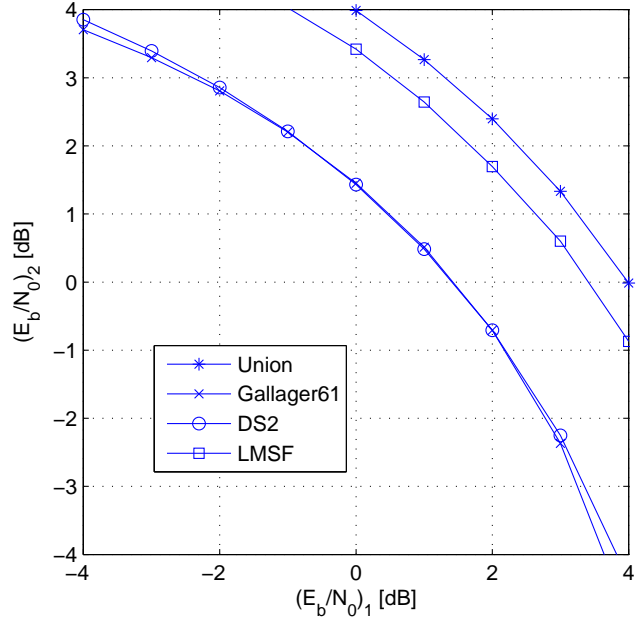

Figure 6: Attainable channel regions for the rate one-third uniformly interleaved ensemble of NSRA$(N, 3)$ codes (see Fig. 5 (a)) in the asymptotic case where we let $N$ tend to infinity. The communication takes place over $J = 2$ parallel binary-input AWGN channels, and the bits are equally likely to be assigned over one of these channels ($\alpha_1 = \alpha_2 = \frac{1}{2}$). The achievable channel region refers to optimal ML decoding. The boundaries of the union and LMSF bounds refer to the discussion in [21], while the boundaries referring to the DS2 bound and the 1961 Gallager bound refer to the derivation in Sections 3 and 4, followed by an optimization of the tilting measures derived in these sections.

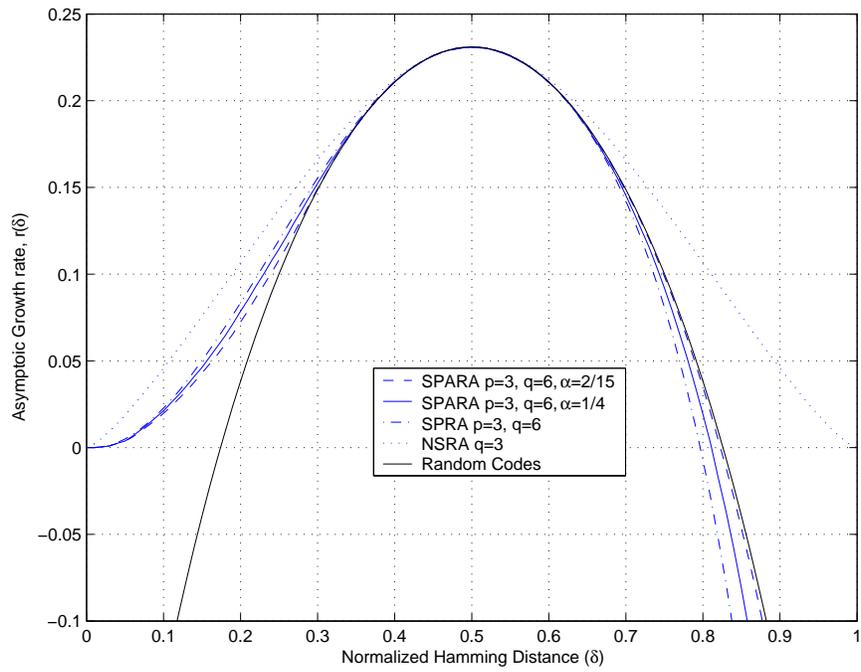

Figure 7: Comparison of asymptotic growth rates of the average distance spectra of ensembles of RA and ARA codes.

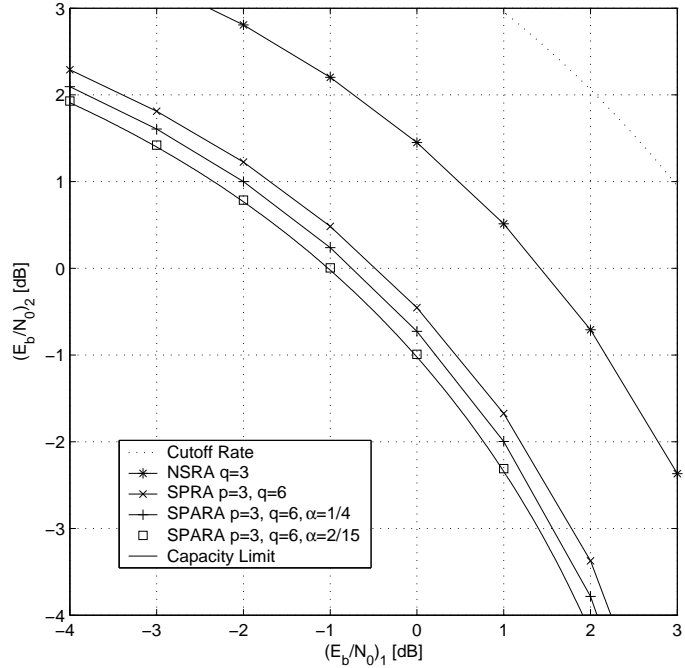

Figure 8: Attainable channel regions for the rate one-third uniformly interleaved accumulate-based ensembles with puncturing depicted in Fig. 5. These regions refer to the asymptotic case where we let $N$ tend to infinity. The communication takes place over $J = 2$ parallel binary-input AWGN channels, and the bits are equally likely to be assigned over one of these channels ($\alpha_1 = \alpha_2 = \frac{1}{2}$). The achievable channel region refers to optimal ML decoding. The boundaries of these regions are calculated via the generalization of the 1961 Gallager bound followed by the optimization of the tilting measures (see Section 4). The capacity limit and the attainable channel regions which corresponds to the cutoff rate are given as a reference.

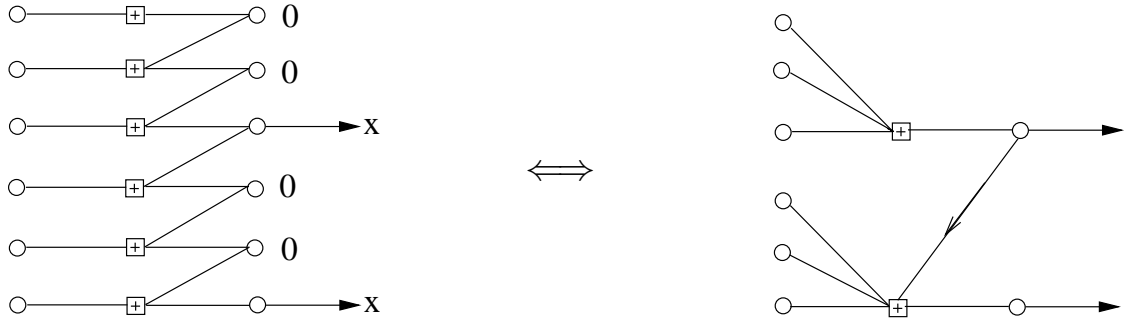

Figure 9: Accumulate code with puncturing period $p = 3$ and an equivalent version of an SPC($p$) code followed by an accumulate code.